\documentclass[lettersize,journal]{IEEEtran}

\usepackage{amsmath,amsfonts}
\usepackage{textcomp}
\usepackage{stfloats}
\usepackage{url}
\usepackage{verbatim}
\usepackage[pdftex]{graphicx}
\def\BibTeX{{\rm B\kern-.05em{\sc i\kern-.025em b}\kern-.08em
    T\kern-.1667em\lower.7ex\hbox{E}\kern-.125emX}}
\usepackage{balance}

\usepackage{tabularx}
\usepackage{subcaption}
\usepackage{color, colortbl}
\definecolor{LightGray}{gray}{0.7}
\usepackage{multirow}

\usepackage[colorinlistoftodos]{todonotes} % should be removed later

\usepackage{tikz}
\tikzset{>=latex}
\usetikzlibrary{arrows,shapes,positioning,calc,math,fadings,patterns,decorations}
\usetikzlibrary{decorations.pathreplacing,decorations.pathmorphing}			
\usetikzlibrary{arrows,arrows.meta}
\usetikzlibrary{shapes.geometric}

\usepackage[acronym]{glossaries}
\setacronymstyle{long-short}
\newacronym{rf}{RF}{radio frequency}
\newacronym{uav}{UAV}{unmanned aerial vehicle}
\newacronym{cnn}{CNN}{convolutional neural network}
\newacronym{usrp}{USRP}{universal software radio peripheral}
\newacronym{sdr}{SDR}{software-defined radio}
\newacronym{iq}{IQ}{in-phase and quadrature}
\newacronym{rms}{RMS}{root mean square}
\newacronym{snr}{SNR}{signal-to-noise ratio}
\newacronym{ml}{ML}{machine learning}
\newacronym{dl}{DL}{deep learning}
\newacronym{gradcam}{Grad-CAM}{Gradient-weighted class activation mapping}
\newacronym{vgg}{VGG}{Visual Geometry Group}
\newacronym{tsne}{t-SNE}{t-distributed Stochastic Neighbor Embedding}
\newacronym{umap}{UMAP}{Uniform Manifold Approximation and Projection}

\begin{document}
\title{Robust Low-Cost Drone Detection and Classification Using Convolutional Neural Networks in Low SNR Environments}
\author{Stefan Glüge, Matthias Nyfeler, Ahmad Aghaebrahimian, Nicola Ramagnano, Christof Schüpbach %\IEEEmembership{Life~Fellow,~IEEE}

\thanks{S. Glüge, M. Nyfeler, A. Aghaebrahimian are with the Institute of Computational Life Sciences, Zurich University of Applied Sciences, Wädenswil, Switzerland}
\thanks{N. Ramagnano is with the Institute for Communication Systems, Eastern Switzerland University of Applied Sciences, Rapperswil-Jona, Switzerland}
\thanks{C. Schüpbach is with Armasuisse Science + Technology, Thun, Switzerland}}

\markboth{IEEE Journal of Radio Frequency Identification}%
{S. Glüge \MakeLowercase{\textit{et al.}}: Robust Low-Cost Drone Detection and Classification Using Convolutional Neural Networks in Low SNR Environments}

\maketitle

\begin{abstract}
The proliferation of drones, or \glspl{uav}, has raised significant safety concerns due to their potential misuse in activities such as espionage, smuggling, and infrastructure disruption. This paper addresses the critical need for effective drone detection and classification systems that operate independently of \gls{uav} cooperation. We evaluate various \glspl{cnn} for their ability to detect and classify drones using spectrogram data derived from consecutive Fourier transforms of signal components. The focus is on model robustness in low \gls{snr} environments, which is critical for real-world applications. A comprehensive dataset is provided to support future model development. In addition, we demonstrate a low-cost drone detection system using a standard computer, \gls{sdr} and antenna, validated through real-world field testing. On our development dataset, all models consistently achieved an average balanced classification accuracy of $\ge 85\%$ at \gls{snr} $>-12$\,dB. In the field test, these models achieved an average balance accuracy of $>80\%$, depending on transmitter distance and antenna direction. Our contributions include: a publicly available dataset for model development, a comparative analysis of \gls{cnn} for drone detection under low \gls{snr} conditions, and the deployment and field evaluation of a practical, low-cost detection system. 
\end{abstract}


\section{Introduction}
\label{sec:intro}
% Every manuscript must:
% 1. provide a clear statement of the problem and what the contribution of the work is to the relevant research community;
% 2. state why this contribution is significant (what impact it will have);
% 3. provide citation of the published literature most closely related to the manuscript; and
% 4. state what is distinctive and new about the current manuscript relative to these previously published works.

% Manuscript Length:
% 1. For the initial submission of a regular paper, the manuscript may not exceed 13 double-column pages (10 point font), including title; names of authors and their complete contact information; abstract; text; all images, figures and tables, appendices and proofs; and all references. Supplemental materials and graphical abstracts are not included in the page count. For the IEEE Transactions on Multimedia, the initial submission of a regular paper may not exceed 10 pages.
% 2. For regular papers, the revised manuscript may not exceed 16 double-column pages (10 point font), including title; names of authors and their complete contact information; abstract; text; all images, figures and tables, appendices and proofs; and all references.

\IEEEPARstart{D}{rones}, or civil \glspl{uav}, have evolved from hobby toys to commercial systems with many applications. In particular, mini/amateur drones have become ubiquitous. With the proliferation of these low-cost, small and easy-to-fly drones, safety issues have became more pressing (e.g.\ spying, transfer of illegal or dangerous goods, disruption of infrastructure, 
assault). Although regulations and technical solutions (such as transponder systems) are in place to safely integrate \glspl{uav} into the airspace, detection and classification systems that do not rely on the cooperation of the \gls{uav} are necessary. Various technologies such as audio, video, radar, or \gls{rf} scanners have been proposed for this task \cite{Al-lQubaydhi2024}.

In this paper, we evaluate different \glspl{cnn} for drone detection and classification using the spectrogram data computed with consecutive Fourier transforms for the real and imaginary parts of the signal. To facilitate future model development, we make the dataset publicly available. In terms of performance, we focus on the robustness of the models to low \glspl{snr}, as this is the most relevant aspect for a real-world application of the system. Furthermore, we evaluate a low-cost drone detection system consisting of a standard computer, \gls{sdr}, and antenna in a real-world field test. 

Our contributions can therefore be summarised as follows:
\begin{itemize}
    \item We provide the dataset used to develop the model. Together with the code to load and transform the data, it can be easily used for future model development.
    \item We compare different \glspl{cnn} using $2$D spectrogram data for detection and classification of drones based on their \gls{rf} signals under challenging conditions, i.e.\ low \glspl{snr} down to $-20$\,dB.
    \item We visualise the model embeddings to understand how the model clusters and separates different classes, to identify potential overlaps or ambiguities, and to examine the hierarchical relationships within the learned features.  
    \item We implement the models in a low-cost detection system and evaluate them in a field test.
\end{itemize}

\subsection{Related Work}
A literature review on drone detection methods based on \gls{dl} is given in \cite{Al-lQubaydhi2024} and \cite{Rahman2024}. Both works reflect the state of the art in $2024$. Different \gls{dl} algorithms are discussed with respect to the techniques used to detect drones based on visual, radar, acoustic, and \gls{rf} signals. Given these general overviews, we briefly summarise recent work based on \gls{rf} data, with a particular focus on the data side of the problem to motivate our work.

With the advent of \gls{dl}-based methods, the data used to train models became the cornerstone of any detection system. Table \ref{tab:drone_rf_datasets} provides an overview of openly available datasets of \gls{rf} drone signals. The DroneRF dataset \cite{Allahham2019_DroneRF_dataset} is one of the first openly available datasets. It contains \gls{rf} time series data from three drones in four flight modes (i.e.\ on, hovering, flying, video recording) recorded by two \gls{usrp} \gls{sdr} transceivers \cite{Al-Sad2019_DroneRF_dataset_article}. The dataset is widely used and enabled follow-up work with different approaches to classification systems, i.e.\ \gls{dl}-based \cite{Swinney2020, Zhang2021}, focused on pre-processing and combining signals from two frequency bands \cite{Ge2021}, genetic algorithm-based heterogeneous integrated k-nearest neighbour \cite{Xue2024}, and hierarchical reinforcement learning-based \cite{AlKhonaini2024}. In general, the classification accuracies reported in the papers on the DroneRF dataset are close to $100\%$. Specifically, \cite{Al-Sad2019_DroneRF_dataset_article}, \cite{Swinney2020}, and \cite{Zhang2021} report an average accuracy of $99.7\%$, $100\%$, and $99.98\%$, respectively, to detect the presence of a drone. There is therefore an obvious need for a harder, more  realistic dataset. %Signals that could be considered noise in the $2.4$\,GHz band, such as Bluetooth and Wi-Fi, were not recorded. 

Consequently, \cite{Ezuma2020} investigate the detection and classification of drones in the presence of Bluetooth and Wi-Fi signals. Their system used a multi-stage detector to distinguish drone signals from the background noise and interfering signals. Once a signal was identified as a drone signal, it was classified using \gls{ml} techniques. The detection performance of the proposed system was evaluated for different \glspl{snr}. The corresponding recordings ($17$ drone controls from eight different manufacturers) are openly available \cite{Ezuma2020data_drone_remote_controller}. Unfortunately, the Bluetooth/Wi-Fi noise is not part of the dataset. Ozturk et al.\ \cite{Ozturk2021} used the dataset to further investigate the classification of \gls{rf} fingerprints at low \glspl{snr} by adding white Gaussian noise to the raw data. Using a \gls{cnn}, they achieved classification accuracies ranging from $92\%$ to $100\%$ for \gls{snr} $\in[-10, 30]$dB.

The openly available DroneDetect dataset \cite{Sweeny2021data_drone_detect} was created by Swinney and Woods \cite{Swinney2021}. It contains raw \gls{iq} data recorded with a BladeRF \gls{sdr}. Seven drone models were recorded in three different flight modes (on, hovering, flying). Measurements were also repeated with different types of noise, such as interference from a Bluetooth speaker, a Wi-Fi hotspot, and simultaneous Bluetooth and Wi-Fi interference. The dataset does not include measurements without drones, which would be necessary to evaluate a drone detection system. The results in \cite{Swinney2021} show that Bluetooth signals are more likely to interfere with detection and classification accuracy than Wi-Fi signals. Overall, frequency domain features extracted from a \gls{cnn} were shown to be more robust than time domain features in the presence of interference. In \cite{Kunze2022_drone_cnn_raw_iq_data} the drone signals from the DroneDetect dataset were augmented with Gaussian noise and \gls{sdr} recorded background noise. Hence, the proposed approach could be evaluated regrading its capability to detect drones. They trained a \gls{cnn} end-to-end on the raw \gls{iq} data and report an accuracy of $99\%$ for detection and between $72\%$ and $94\%$ for classification.

The Cardinal RF dataset \cite{Medaiyese2022_data_cardRF} consists of the raw time series data from six drones + controller, two Wi-Fi and two Bluetooth devices. Based on this dataset, Medaiyese et al.~\cite{Medaiyese2021_semi_sup_framework} proposed a semi-supervised framework for \gls{uav} detection using wavelet analysis. Accuracy between $86\%$ and $97\%$ was achieved at \glspl{snr} of $30$\,dB and $18$\,dB, while it dropped to chance level for \glspl{snr} below $10$\,dB to $6$\,dB. In addition, \cite{Medaiyese2022} investigated different wavelet transforms for the feature extraction from the \gls{rf} signals. Using the wavelet scattering transform from the steady state of the \gls{rf} signals at $30$\,dB \gls{snr} to train SqueezeNet \cite{Iandola2016_squeezenet}, they achieved an accuracy of $98.9\%$ at $10$\,dB \gls{snr}.

In our previous work \cite{Gluege2023}, we created the noisy drone \gls{rf} signals dataset\footnote{\url{https://www.kaggle.com/datasets/sgluege/noisy-drone-rf-signal-classification}} from six drones and four remote controllers. It consists of non-overlapping signal vectors of $16384$ samples, corresponding to $\approx1.2$\,ms at $14$\,MHz. We added Labnoise (Bluetooth, Wi-Fi, Amplifier) and
Gaussian noise to the dataset and mixed it with the drone signals with \gls{snr} $\in[-20, 30]$\,dB. Using \gls{iq} data and spectrogram data to train different \glspl{cnn}, we found an advantage in favour of the $2$D spectrogram representation of the data. There was no performance difference at \gls{snr} $\ge0$\,dB but a major improvement in the balanced accuracy at low \gls{snr} levels, i.e.\ $84.2$\% on the spectrogram data compared to $41.3$\% on the \gls{iq} data at $-12$\,dB \gls{snr}.

Recently, \cite{Zhao2024} proposed an anchor-free object detector based on keypoints for drone \gls{rf} signal spectograms. They also proposed an adversarial learning-based data adaptation method to generate domain independent and domain aligned features. Given five different types of drones, they report a mean average precision of $97.36\%$, which drops to $\approx 55\%$ when adding Gaussian noise with $-25$\,dB \gls{snr}. The raw data used in their work is available\footnote{\url{https://www.kaggle.com/datasets/zhaoericry/drone-rf-dataset}}, but yet, unfortunately not usable without any further documentation.

% Zhang et al.~\cite{Zhang2023} used downlink video data and focused on data augmentation with various environmental signals. They applied data segmentation techniques of spectral features with a ResNet \cite{He2016_resnet} architecture for the \gls{rf} data with a bandwidth of $100$\,MHz with \glspl{snr} ranging from $20$\,dB down to $0$\,dB. The accuracy at the lowest \gls{snr} of $0$\,dB was still around $70\%$.

\begin{table*}
\newcolumntype{C}{>{\centering\scriptsize\arraybackslash}X}
\newcolumntype{L}{>{\scriptsize\arraybackslash}X}
\newcolumntype{R}{>{\hfill\scriptsize\arraybackslash}X}
\caption{Overview on openly available drone \gls{rf} datasets.}
\begin{tabularx}{\textwidth}{L|CCCCC}
\hline
    \textbf{Dataset} & \textbf{Year} & \textbf{Datatype} & \textbf{\gls{uav}} & \textbf{Noise} & \textbf{Size} \\
\hline    
    DroneRF \cite{Allahham2019_DroneRF_dataset} & 2019 & Raw Amplitude & $3$ drones + $3$ controller & Background RF activities & $3.75$\,GB\\
    Drone remote controller RF signal dataset \cite{Ezuma2020data_drone_remote_controller} & 2020 & Raw Amplitude & $17$ controller & none & $124$\,GB\\
    DroneDetect dataset \cite{Sweeny2021data_drone_detect} & 2020 & Raw IQ & $7$ drones + $7$ controller & Bluetooth, Wi-Fi devices& $66$\,GB\\
    Cardinal RF \cite{Medaiyese2022_data_cardRF} & 2022 & Raw Amplitude & $6$ drones + $6$ controller & Bluetooth, Wi-Fi & $65$\,GB\\
    Noisy drone RF signals \cite{Gluege2023} & 2023 & Pre-processed IQ and Spectrogram & $6$ drones + $4$ controller & Bluetooth, Wi-Fi, Gauss & $23$\,GB\\
\hline
\end{tabularx}
\label{tab:drone_rf_datasets}
\end{table*}

\subsection{Motivation}
As we have seen in other fields, such as computer vision, the success of \gls{dl} can be attributed to: (a) high-capacity models; (b) increased computational power; and (c) the availability of large amounts of labelled data \cite{Sun2017}. Thus, given the large amount of available raw \gls{rf} signals (cf.\ Tab.\ \ref{tab:drone_rf_datasets}) we promote the idea of open and reusable data, to facilitate model development and model comparison.

With the noisy drone \gls{rf} signals dataset \cite{Gluege2023}, we have provided a first ready-to-use dataset to enable rapid model development, without the need for any data preparation. Furthermore, the dataset contains samples that can be considered as ``hard'' in terms of noise, i.e.\ Bluetooth + Wi-Fi + Gaussian noise at very low \glspl{snr}, and allows a direct comparison with the published results.

While the models proposed in \cite{Gluege2023} performed reasonably well in the training/lab setting, we found it difficult to transfer their performance to practical application. The reason was the choice of rather short signal vectors of $16384$ samples, corresponding to $\approx1.2$\,ms at $14$\,MHz. Since the drone signals occur in short bursts of $\approx 1.3 \text{--} 2$\,ms with a repetition period of $\approx 60 \text{--} 600$\,ms, our continuously running classifier predicts a drone whenever a burst occurs and noise during the repetition period of the signal. Therefore, in order to provide a stable and reliable classification per every second, one would need an additional ``layer'' to pool the classifier outputs given every $1.2$\,ms.

In the present work, we follow a data-centric approach and simply increase the length of the input signal to $\approx 75$\,ms to train a classifier in an end-to-end manner. Again, we provide the data used for model development in the hope that it will inspire others to develop better models. 

In the next section, we briefly describe the data collection and preprocessing procedure. Section \ref{sec:methods} describes the model architectures and their training/validation method. In addition, we describe the setup of a low-cost drone detection system and of the field test. The resulting performance metrics are presented in Section \ref{sec:results} and are further discussed in Section \ref{sec:discussion}.

%%%%%%%%%%%%%%%%%%%%%%%%%%%%%%%%%%%%%%%%%%
\section{Materials}
\label{sec:material}
\noindent We used the raw \gls{rf} signals from the drones that were collected in \cite{Gluege2023}. Nevertheless, we briefly describe the data acquisition process again to provide a complete picture of the development from the raw \gls{rf} signal to the deployment of a detection system within a single manuscript.

\subsection{Data Acquisition}
\label{sec:data_collection}
\begin{figure}
\includegraphics[width=\columnwidth]{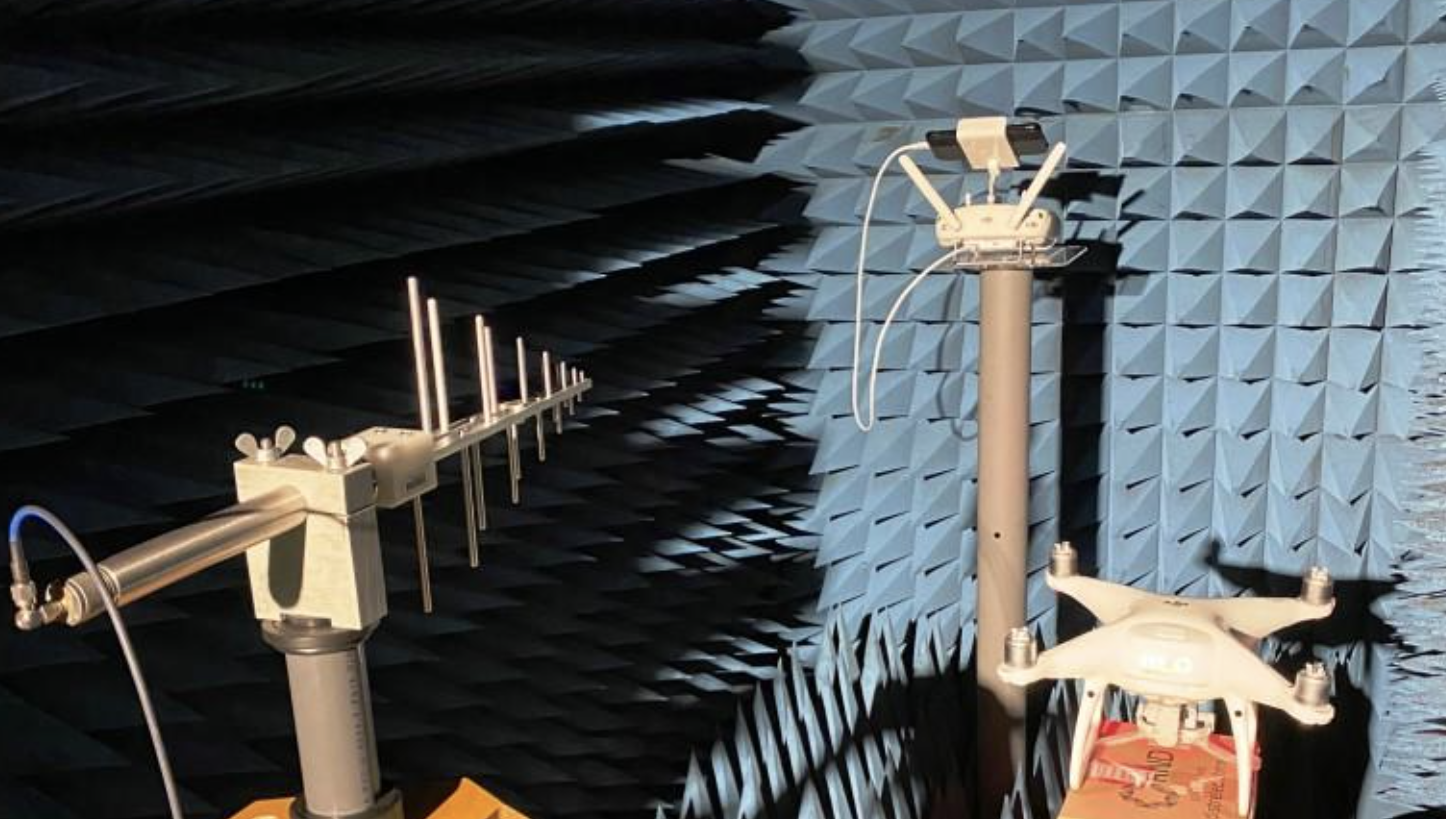}
\caption{Recording of drone signals in the anechoic chamber. A DJI Phantom 4 Pro drone with the DJI Phantom GL300F remote control.}
\label{fig:dev_dataset_recording}
\end{figure}

The drone's remote control and, if present, the drone itself were placed in an anechoic chamber to record the raw \gls{rf} signal without interference for at least one minute. The signals were received by a log-periodic antenna and sampled and stored by an Ettus Research USRP B210, see Fig.~\ref{fig:dev_dataset_recording}. In the static measurement, the respective signals of the remote control (TX) alone or with the drone (RX) were measured. In the dynamic measurement, one person at a time was inside the anechoic chamber and operated the remote control (TX) to generate a signal that is as close to reality as possible. All signals were recorded at a sampling frequency of $56$\,MHz (highest possible real-time bandwidth). All drone models and recording parameters are listed in Tab.~\ref{tab:dev_dataset_recording}, including both uplink and downlink signals.

\begin{table*}
\centering
\caption{Transmitters and receivers recorded in the development dataset and their respective class labels. Additionally, we show the center frequency (GHz), the channel spacing (MHz), the burst duration (ms), and the repetition period of the respective signals (ms).}
\setlength{\tabcolsep}{3pt}
\begin{tabular}{lllcccc}
\hline
\textbf{Transmitter} & \textbf{Receiver} & \textbf{Label} & \textbf{Center Freq. (GHz)} & \textbf{Spacing (MHz)} & \textbf{Duration (ms)} & \textbf{Repetition (ms)}\\
\hline
DJI Phantom GL300F & DJI Phantom 4 Pro & DJI & $2.44175$ & $1.7$ & $2.18$ & $630$\\
Futaba T7C & - & FutabaT7 & $2.44175$ & $2$ & $1.7$ & $288$ \\
Futaba T14SG & Futaba R7008SB & FutabeT14 & $2.44175$ & $3.1$ & $1.4$ & $330$ \\
Graupner mx-16 & Graupner GR-16 & Graupner & $2.44175$ & $1$ & $1.9/3.7$ & $750$ \\
Bluetooth/Wi-Fi Noise & - & Noise & $2.44175$ & & & \\
Taranis ACCST & X8R Receiver & Taranis & $2.440$ & $1.5$ & $3.1/4.4$ & $420$ \\
Turnigy 9X & - & Turnigy & $2.445$ & $2$ & $1.3$ & $61$, $120$-$2900$ $^{\mathrm{a}}$ \\
\hline
\multicolumn{7}{p{\textwidth}}{$^{\mathrm{a}}$ The repetition period of the Turnigy transmitter is not static. First bursts were observed after $61$\,ms, the following signal bursts were observed in the interval $[120, 2900]$\,ms}
\end{tabular}
\label{tab:dev_dataset_recording}
\end{table*}

We also recorded three types of noise and interference. First, Bluetooth/Wi-Fi noise was recorded using the hardware setup described above. Measurements were taken in a public and busy university building. In this open recording setup, we had no control over the exact number or types of active Bluetooth/Wi-Fi devices and the actual traffic in progress. 

Second, artificial white Gaussian noise was used, and third, receiver noise was recorded for $30$ seconds from the \gls{usrp} at various gain settings ($[30,70]$\,db in steps of $10$\,dB) without the antenna attached. This should prevent the final model from misclassifying quantisation noise in the absence of a signal, especially at low gain settings.

\subsection{Data Preparation}
\label{sec:data_preprocessing}
To reduce memory consumption and computational effort, we reduced the bandwidth of the signals by downsampling from $56$\,MHz to $14$\,MHz using the SciPy \cite{2020SciPy-NMeth} signal.decimate function with an $8$th order Chebyshev type I filter. 

The drone signals occur in short bursts with some low power gain or background noise in between (cf.\ Tab.~\ref{tab:dev_dataset_recording}). We divided the signals into non-overlapping vectors of $1048576$ samples ($74.9$\,ms) and only vectors containing a burst, or at least a partial burst, were used for the development dataset. This was achieved by applying an energy threshold. As the recordings were made in an echo-free chamber, the signal burst is always clearly visible. Hence, we only used vectors that contained a portion of the signal whose energy was above the threshold, which was arbitrarily set at $0.001$ of the average energy of the entire recording.

The selected drone signal vectors $x$ with $i\in\{1,\ldots k\}$ were normalised to a carrier power of $1$ per sample, i.e.\ only the part of the signal vector containing drone bursts was considered for the power calculation ($m$ samples out of $k$). This was achieved by identifying the bursts as those samples where a smoothed energy was above a threshold. %, as shown in Figure~\ref{fig:carrier}. 

The signal vectors $x$ are thus normalised by
\begin{equation}
  \hat{x}(i) = x(i)\ / \ \sqrt{\frac{1}{m}\sum_i | x(i) | ^2}.
\end{equation}

Noise vectors (Bluetooth, Wi-Fi, Amplifier, Gauss) $n$ with samples $i\in\{1,\ldots k\}$ were normalised to a mean power of $1$ with 
\begin{equation}
    \hat{n}(i) = n(i) / \sqrt{\frac{1}{k}\sum_i | n(i) | ^2}.
\end{equation}

Finally, the normalised drone signal vectors were mixed with the normalised noise vectors by
%at different \glspl{snr} with% Since the signal carrier power, and the noise power, were both normalized to $1$, we added separate normalized noise vectors $\hat{n}$ to each normalized signal vector $\hat{x}$ as
\begin{equation}
\hat{y}(i) = \frac{\left(\sqrt{k} \cdot \hat{x}(i)+\hat{n}(i)\right)}{\sqrt{k + 1}},\text{ with }k=10^{\frac{\text{SNR}}{10}},
\end{equation}
to generate the noisy drone signal vectors $\hat{y}$ at different \glspl{snr}.

\subsection{Development Dataset}
\label{sec:dev_dataset}
To facilitate future model development, we provide our resulting dataset\footnote{\url{https://www.kaggle.com/datasets/sgluege/noisy-drone-rf-signal-classification-v2}} along with a code example\footnote{\url{https://github.com/sgluege/noisy-drone-rf-signal-classification-v2}} to load and inspect the data. The dataset consists of the non-overlapping signal vectors of $2^{20}$ samples, corresponding to $\approx 74.9$\,ms at $14$\,MHz. 

As described in Sec.~\ref{sec:data_preprocessing}, the drone signals were mixed with noise (cf.~Eqs. $1$-$3$). More specifically, $50\%$ of the drone signals were mixed with Labnoise (Bluetooth + Wi-Fi + Amplifier) and $50\%$ with Gaussian noise. In the same way we created a single noise class by mixing Labnoise and Gaussian noise in all possible combinations (i.e., Labnoise $+$ Labnoise, Labnoise $+$ Gaussian noise, Gaussian noise $+$ Labnoise, and Gaussian noise $+$ Gaussian noise). This mixing was done as for the drone signals using Eq.~$3$. For instance, in case of a Labnoise $+$ Gaussian noise mix, $\hat{x}$ refers to a Labnoise vector and $\hat{n}$ to a generated Gaussian noise vector. 

For the drone signal classes, as for the noise class, the number of samples for each \gls{snr} level was evenly distributed over the interval of \glspl{snr} $\in [-20, 30]$\,dB in steps of $2$\,dB, i.e., $679$-$685$ samples per \gls{snr} level. The resulting number of samples per class is given in Tab.~\ref{tab:class_stats_dev_dataset}.

\begin{table}
\centering
\caption{Number of samples in the different classes in the development dataset.}
\setlength{\tabcolsep}{2pt}
\begin{tabular}{l|ccccccc}
\hline
\textbf{Class} & DJI & FutabaT14 & FutabaT7 & Graupner & Taranis & Turnigy & Noise\\
\textbf{\#samples} & $1280$ & $3472$ & $801$ & $801$ & $1663$ & $855$ & $8872$\\
\hline
\end{tabular}
\label{tab:class_stats_dev_dataset}
\end{table}

\begin{figure*}
\centering
\begin{subfigure}[b]{0.49\textwidth}
\centering
\includegraphics[width=\textwidth]{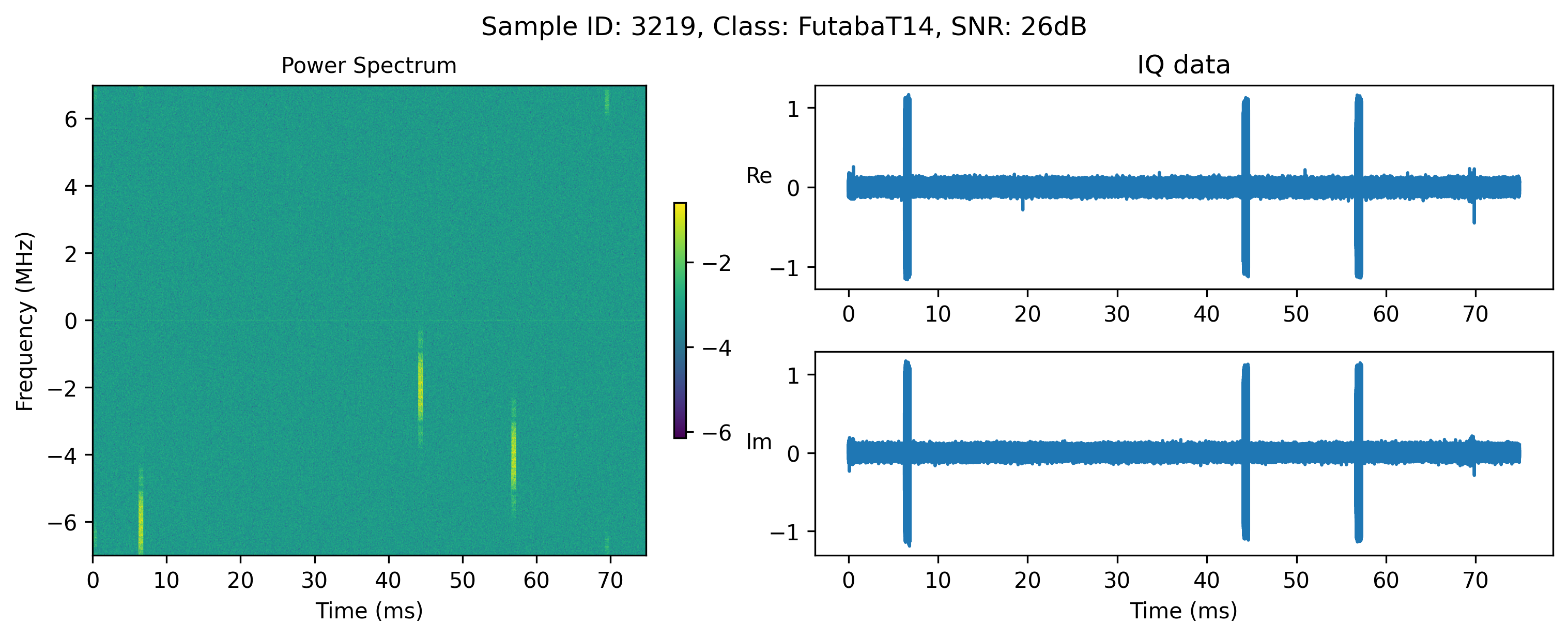}
\caption{FutabaT14 at SNR $26$\,dB}
\label{fig:input_sample_a}
\end{subfigure}
\hfill
\begin{subfigure}[b]{0.49\textwidth}
\centering
\includegraphics[width=\textwidth]{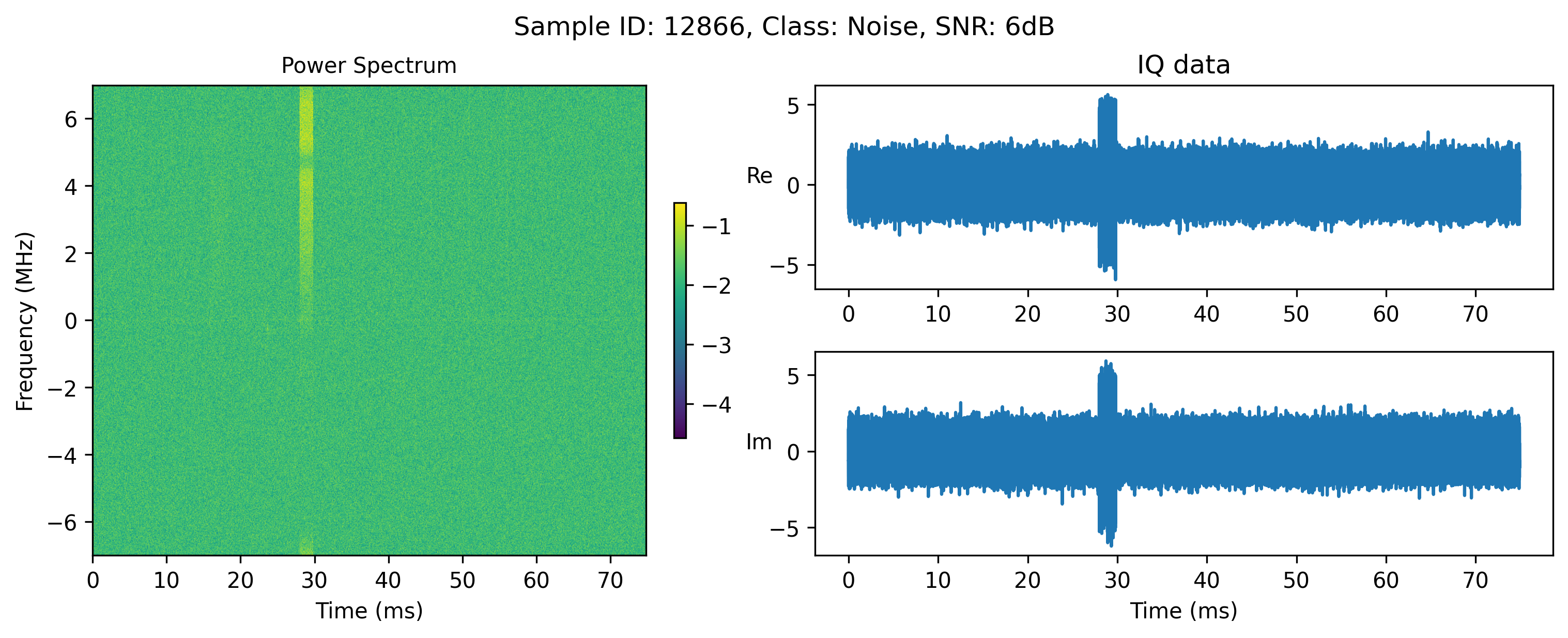}
\caption{DJI at SNR $6$\,dB}
\label{fig:input_sample_b}
\end{subfigure}
\\
\begin{subfigure}[b]{0.49\textwidth}
\centering
\includegraphics[width=\textwidth]{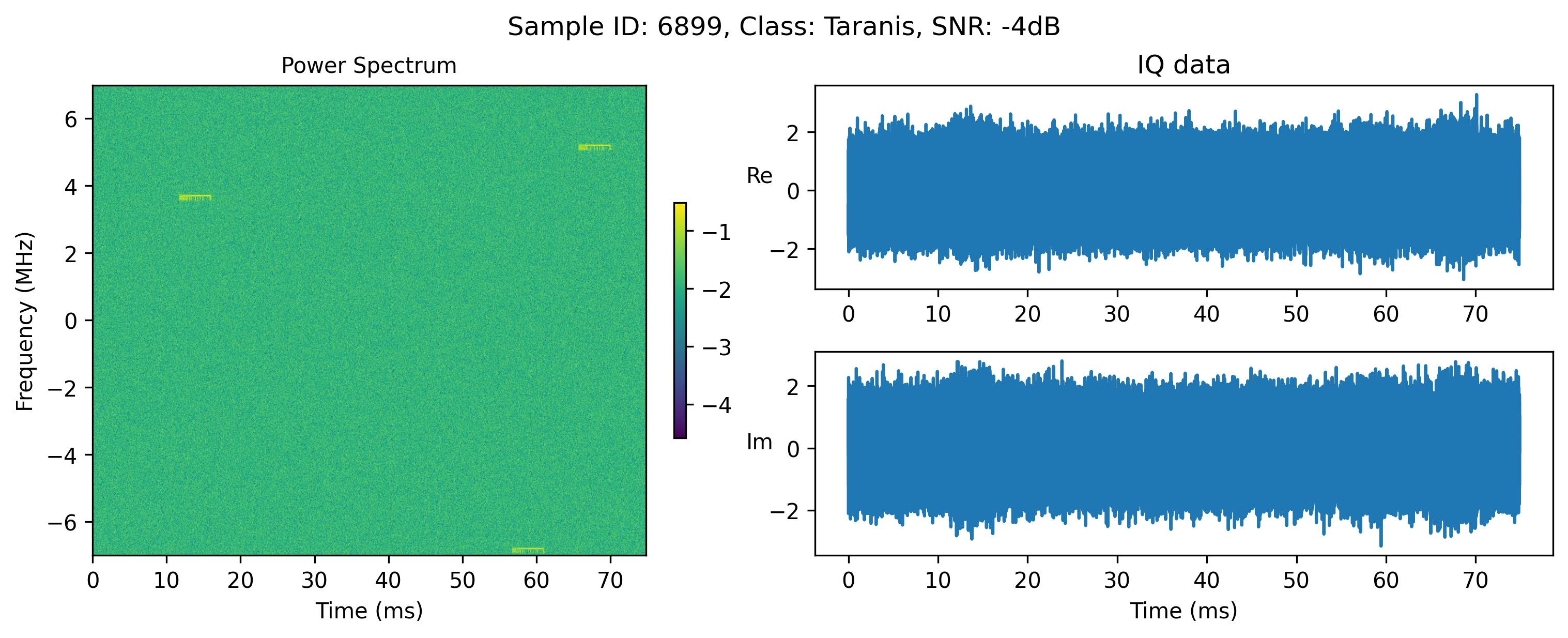}
\caption{Taranais at SNR $-4$\,dB}
\label{fig:input_sample_c}
\end{subfigure}
\hfill
\begin{subfigure}[b]{0.49\textwidth}
\centering
\includegraphics[width=\textwidth]{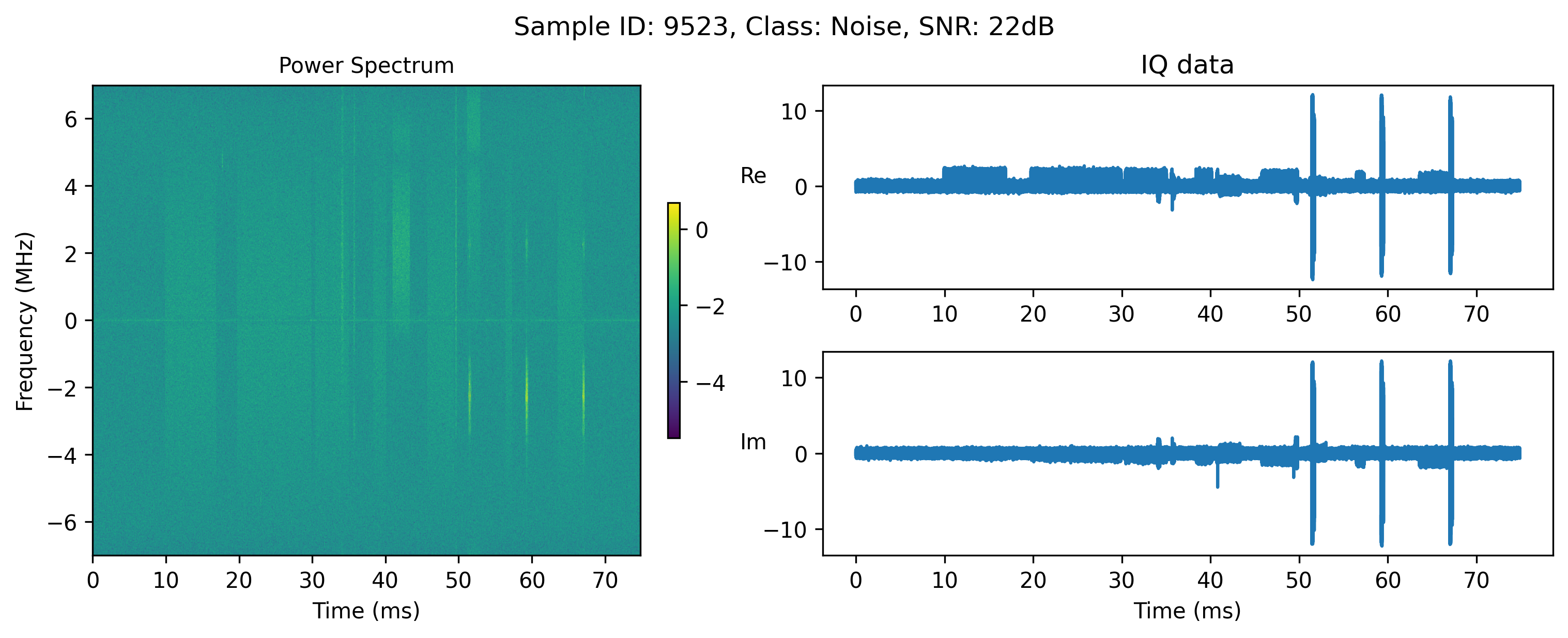}
\caption{Noise at SNR $22$\,dB}
\label{fig:input_sample_d}
\end{subfigure}
\caption{Log power spectrogram and IQ data samples from the development dataset at different \glspl{snr} (\subref{fig:input_sample_a}-\subref{fig:input_sample_d})}
\label{fig:input_samples}
\end{figure*}

In our previous work \cite{Gluege2023} we found an advantage in using the spectrogram representation of the data compared to the \gls{iq} representation, especially at low \glspl{snr} levels. Therefore, we transform the raw \gls{iq} signals by computing the spectrum of each sample with consecutive Fourier transforms with non-overlapping segments of length $1024$ for the real and imaginary parts of the signal. That is, the two \gls{iq} signal vectors ($[2 \times 2^{20}]$) are represented as two matrices ($[2 \times 1024 \times 1024]$). Fig.~\ref{fig:input_samples} shows four samples of the dataset at different \glspl{snr}. Note that we have plotted the log power spectrogram of the complex spectrum $\hat{y}_{\mathrm{fft}}$ as
\begin{equation}
  \log_{10}|\hat{y}_{\mathrm{fft}}| = \log_{10}(\sqrt{\mathrm{Re}(\hat{y}_{\mathrm{fft}})^2 + \mathrm{Im}(\hat{y}_{\mathrm{fft}})^2})
\end{equation}

\subsection{Detection System Prototype}
\label{sec:detection_system_prototype}

\newcommand{\bpf}[4]{
	% 1: x
	% 2: y
	% 3: diameter
	% 4: node name
	\node(#4)[draw, shape=rectangle,inner sep=0, minimum height=#3 cm, minimum width=#3 cm] at (#1,#2){};
	\draw (#1,#2+0.2*#3) ..controls(#1+0.1*#3,#2-0.08*#3+0.2*#3) and (#1+.2*#3,#2-0.08*#3+0.2*#3) .. (#1+.3*#3,#2+0.2*#3);
	\draw (#1,#2+0.2*#3) ..controls(#1-0.1*#3,#2+0.08*#3+0.2*#3) and (#1-.2*#3,#2+0.08*#3+0.2*#3) .. (#1-.3*#3,#2+0.2*#3);
	\draw (#1,#2) ..controls(#1+0.1*#3,#2-0.08*#3) and (#1+.2*#3,#2-0.08*#3) .. (#1+.3*#3,#2);
	\draw (#1,#2) ..controls(#1-0.1*#3,#2+0.08*#3) and (#1-.2*#3,#2+0.08*#3) .. (#1-.3*#3,#2);
	\draw (#1,#2-0.2*#3) ..controls(#1+0.1*#3,#2-0.08*#3-0.2*#3) and (#1+.2*#3,#2-0.08*#3-0.2*#3) .. (#1+.3*#3,#2-0.2*#3);
	\draw (#1,#2-0.2*#3) ..controls(#1-0.1*#3,#2+0.08*#3-0.2*#3) and (#1-.2*#3,#2+0.08*#3-0.2*#3) .. (#1-.3*#3,#2-0.2*#3);
	\draw (#1-0.08*#3,#2-0.08*#3+0.2*#3) -- (#1+0.08*#3,#2+0.08*#3+0.2*#3);
	\draw (#1-0.08*#3,#2-0.08*#3-0.2*#3) -- (#1+0.08*#3,#2+0.08*#3-0.2*#3);
}

\newcommand{\antenna}[4]{
	% 1: x
	% 2: y
	% 3: size
	% 4: node name
    \node[inner sep=0, minimum size=0](#4) at (#1,#2+0.5*#3){};
    \draw (#4) -- ++(0,0.5*#3) -- +(-0.3*#3,0) -- +(0,-0.5*#3) -- +(0.3*#3,0) -- +(-0.3*#3,0);
}

\begin{figure}[!ht]
\centering
\resizebox{\linewidth}{!}{
\begin{tikzpicture}\sffamily\small
\node[draw, minimum height=1cm] (gnuradio) at (0,0){GNU Radio};
\node[draw, minimum height=1cm] (gui) at (2.5,0){GUI};
\node[draw, minimum height=1cm] (pytorch) at (0,1.75){PyTorch};
\node[draw, minimum height=1cm] (gpu) at (2.5,1.75){GPU};
\node[draw, minimum height=1cm, align=center] (usrp) at (-2.75,0){USRP\\B210};
%\node[draw, thick, minimum height=1cm, align=center] (bpf) at (-5.5,0){Band-Pass\\Filter};
%\node[draw, thick, minimum height=1cm, align=center] (ant) at (-5.5,2){2.4\,GHz\\Antenna};
\antenna{-5.5}{0}{1}{ant};
\bpf{-4.4}{0}{1}{bpf};
\node[above] at (bpf.north){\scriptsize 2.4-2.48\,GHz};
\draw[dashed] (-1.5,-1) rectangle (3.5,3);
\node[below right] at (-1.5,3){Mobile computer};

% connections
\draw[->] (ant) |- (bpf);
\draw[->] (bpf) -- (usrp);
\draw[<->] (usrp) --node[above, inner sep=1.5pt, yshift=1.5pt, fill=white]{\scriptsize USB3} (gnuradio);
\draw[<->] (gnuradio) -- (gui);
\draw[<->] (pytorch) -- (gpu);
\draw[->] (gnuradio) to[out=100, in=260]node[left]{\scriptsize IQ Data} (pytorch);
\draw[->] (pytorch) to[out=280, in=80]node[right,align=left]{\scriptsize Classification\\[-0.5ex]\scriptsize Results} (gnuradio);
\end{tikzpicture}}
\caption{Block diagram of the mobile drone detection system.}
\label{fig:block_diagram_mobile_system}
\end{figure}
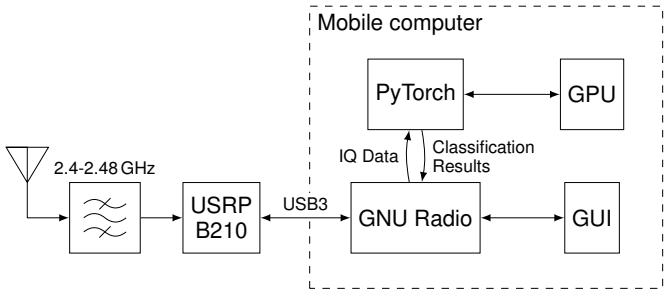

\begin{figure}
\includegraphics[width=\columnwidth]{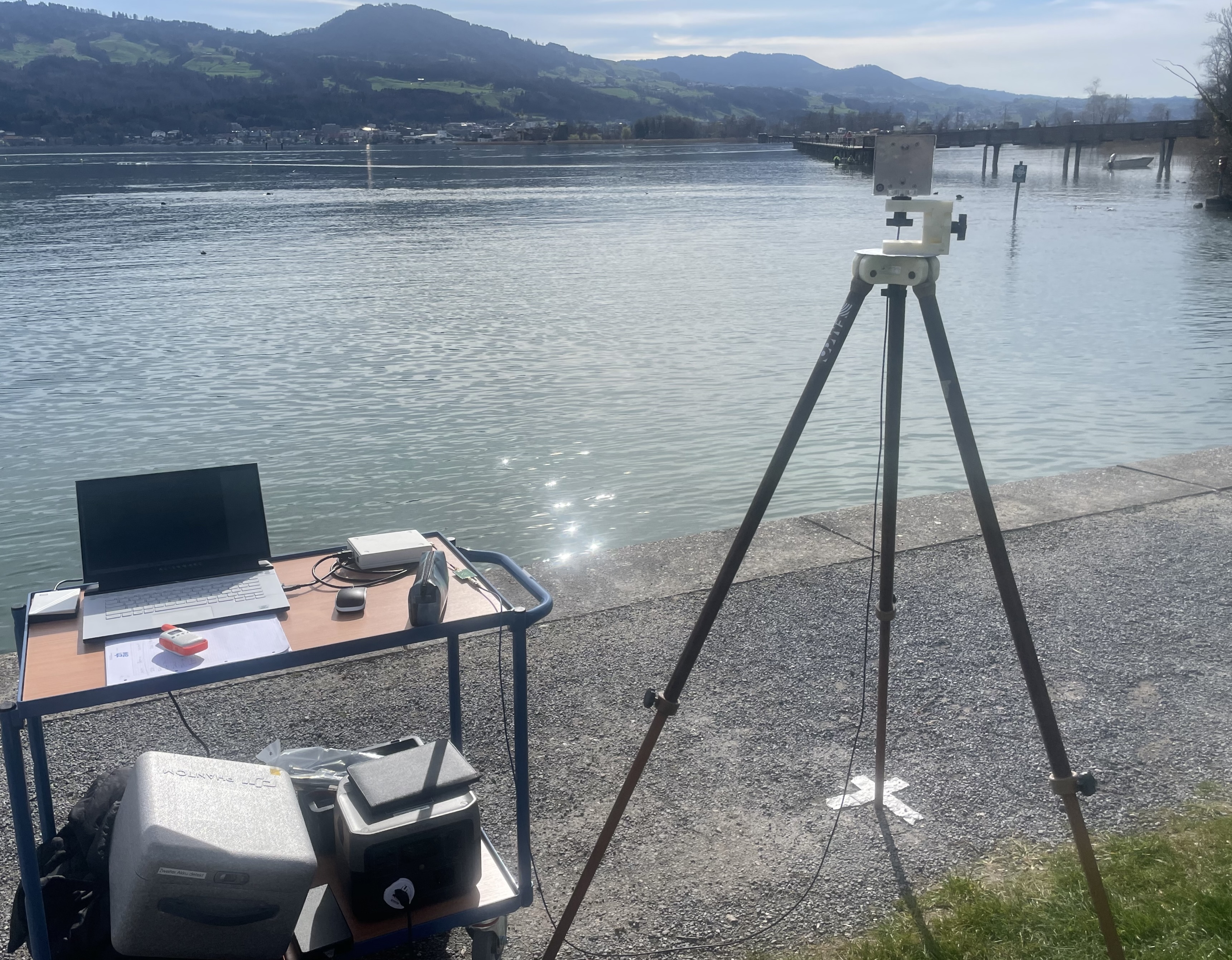}
\caption{Detection prototype at the Zurich Lake in Rapperswil.}
\label{fig:field_test_messstation}
\end{figure}

For field use, a system based on a mobile computer was used as shown in Fig.~\ref{fig:block_diagram_mobile_system} and illustrated in Fig.~\ref{fig:field_test_messstation}. The \gls{rf} signals were received using a directional left-hand circularly polarised antenna (H\&S SPA 2400/70/9/0/CP).
The antenna gain of $8.5$\,dBi and the front-to-back ratio of $20$\,dB helped to increase the detection range and to attenuate the unwanted interferers in the opposite direction. Circular polarisation has been chosen to eliminate the alignment problem as the transmitting antennas have a linear polarisation. The \gls{usrp} B210 was used to down-convert and digitise the \gls{rf} signal at a sampling rate of $14$\,Msps. On the mobile computer, the GNU Radio program collected the baseband \gls{iq} samples in batches of one second and send one batch at a time to our PyTorch model, which classified the signal. To speed up the computations in the model we utilised an Nvidia GPU in computer. The classification results were then visualised in real time in a dedicated GUI.

%%%%%%%%%%%%%%%%%%%%%%%%%%%%%%%%%%%%%%%%%%
\section{Methods}
\label{sec:methods}
\subsection{Model Architecture and Training}
\label{sec:model_training}
As in \cite{Gluege2023} we chose the \gls{vgg} \gls{cnn} architecture \cite{Simonyan2015_vgg} due to its performance history across a wide range of tasks, including medical image classification \cite{Shin2016}, as backbone for object detection \cite{Shaoqing2015}, and audio signal classification \cite{Hershey2017}.

The main idea of this architecture is to use multiple layers of small ($3 \times 3$) convolutional filters instead of larger ones. This is intended to increase the depth and expressiveness of the network, while reducing the number of parameters. There are several variants of this architecture, which differ in the number of convolutional layers ($11$ and $19$, respectively). We used a variant with a batch normalisation \cite{Ioffe2015_batchnorm} layer after the convolutions, denoted as VGG11\_BN to VGG19\_BN. For the dense classification layer, we used $256$ linear units followed by $7$ linear units at the output (one unit per class).

A stratified $5$-fold train-validation-test split was used as follows. In each fold, we trained a network using $80\%$ and $20\%$ of the available samples of each class for training and testing, respectively. Repeating the stratified split five times ensures that each sample was in the test set once in each experiment. Within the training set, $20\%$ of the samples were used as the validation set during training. 

Model training was performed for $200$ epochs with a batch size of $8$. The PyTorch \cite{Paszke2019pytorch} implementation of the Adam algorithm \cite{Diederik2015_adam_optimizer} was used with a learning rate of $0.005$, betas $(0.9, 0.999)$ and weight decay of $0$. It was conducted on a NVIDIA A100 $80$GB GPU, requiring $25$GB of memory and $20$h of computational time for the VGG11\_BN training, and $40$GB of memory and $40$h of computational time for the VGG19\_BN.

\subsection{Model Evaluation}
\label{sec:model_evaluation}
During training, the model was evaluated on the validation set after each epoch. If the balanced accuracy on the validation set increased, it was saved. After training, the model with the highest balanced accuracy on the validation set was evaluated on the withheld test data. The performance of the models on the test data was accessed in terms of classification accuracy and balanced accuracy.

As accuracy simply measures the proportion of correct predictions out of the total number of observations, it can be misleading for unbalanced datasets. In our case, the noise class is over-represented in the dataset (cf.\ Tab.~\ref{tab:class_stats_dev_dataset}). Therefor, we also report the balanced accuracy, which is defined as the average of the recall obtained for each class, i.e.\ it gives equal weight to each class regardless of how frequent or rare it is.

\subsection{Visualisation of Model Embeddings}
Despite their effectiveness, \glspl{cnn} are often criticised for being ``black boxes''. Understanding the feature representations, or embeddings, learned by the \gls{cnn} helps to demystify these models and provide some understanding of their capabilities and limitations.
In general, embeddings are high-dimensional vectors generated by the intermediate layers that capture essential patterns from the input data.

In our case, we chose the least complex VGG11\_BN model to visualise its embeddings. When inferencing the test data, we collected the activations at the last dense classification layer, which consists of $256$ units. Given $3549$ test samples, this results in a $256 \times 3549$ matrix. Using \gls{tsne}\cite{vandermaaten2008_tsne} and \gls{umap} \cite{McInnes2018_umap} as dimensionality reduction techniques, we project these high-dimensional embeddings into a lower-dimensional space, creating interpretable visualisations that reveal the model's internal data representations.

Our goals were to understand how the model clusters and separates different classes, to identify potential overlaps or ambiguities, and to examine the hierarchical relationships within the learned features. 

\subsection{Detection System Field Test}
\label{sec:detection_system_field_test}
\begin{figure}
\centering
\includegraphics[width=\columnwidth]{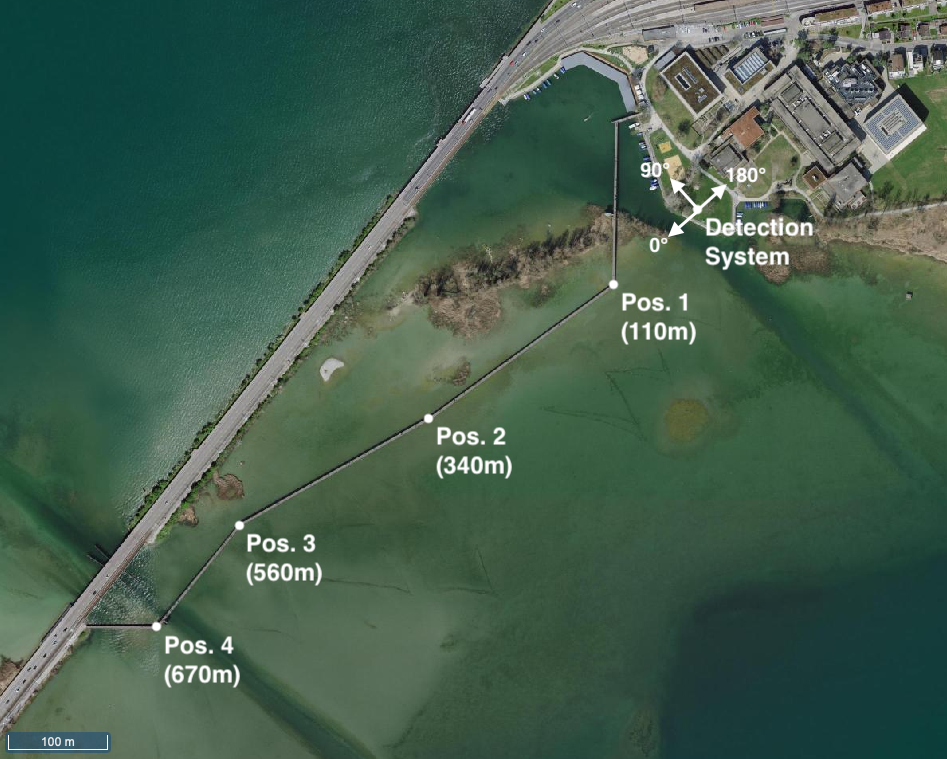}
\caption{Experimental measurement setup at the Zurich Lake in Rapperswil. One can see the four recording positions along the wooden walkway and the detection system positioned at the lake side. Further, recordings were done at different angels of the directional antenna indicated by the arrows at the detection system.}
\label{fig:field_test_sat}
\end{figure}

We conducted a field test of the detection system in Rapperswil at the Zurich Lake. The drone detection prototype was placed on the shore (cf.\ Fig.~\ref{fig:field_test_messstation}) in line of sight of a wooden boardwalk across the lake, with no buildings to interfere with the signals. The transmitters were mounted on a $2.5$\,m long wooden pole. The signals from the transmitters were recorded (and classified in real time) at four positions along the walkway at approximately $110$\,m, $340$\,m, $560$\,m and $670$\,m from the detection system. Figure~\ref{fig:field_test_sat} shows an overview of the experimental setup.
At each recording position, we measured with the directional antenna at three different angles, i.e.\ at $0^{\circ}$ -- facing the drones and/or remote controls, at $90^{\circ}$ -- perpendicular to the direction of the transmitters, and at $180^{\circ}$ -- in the opposite direction. Directing the antenna in the opposite direction should result in $\approx 20$\,dB attenuation of the radio signals.

Table~\ref{tab:field_test_drones_experiment} lists the drones and/or remote controls used in the field test. Note that the Graupner drone and remote control are part of the development dataset (cf.\ Tab.~\ref{tab:dev_dataset_recording}), but were not measured in the field experiment. We assume that no other drones were present during the measurements, so recordings where none of our transmitters were used are labelled as ``Noise''.

\begin{table}
\centering
\caption{Drones and/or remotes used in the field test}
\label{tab:field_test_drones_experiment}
\begin{tabular}{ll}
\hline
\textbf{Class} & \textbf{Drone/remote control} \\
\hline
DJI & DJI Phantom Pro 4 drone and remote\\
FutabaT14 & Futaba T14 remote control\\
FutabaT7 & Futaba T14 remote control\\
Taranis & FrySky Taranis Q X7 remote control\\
Turnigy & Turnigy Evolution remote control\\
\hline
\end{tabular}
\end{table}

For each transmitter, distance, and angle, $20$ to $30$\,s, or approximately $300$ spectrograms were live classified and recorded. The resulting number of samples for each class, distance, and angle are shown in Tab.~\ref{tab:field_test_frequencies}.

\begin{table}
\centering
\caption{Number of samples (\#samples) for each class, distance and antenna direction (angle) recorded in the field test. Recordings at $0$\,m distance have no active transmitter and were therefore labelled “Noise”}
\setlength{\tabcolsep}{3pt}
\begin{tabular}{lc|lc|lc}
  \hline
  \textbf{class} & \textbf{\#samples} & \textbf{Distance [m]} & \textbf{\#samples} & \textbf{Angle [$^{\circ}$]} & \textbf{\#samples}\\
  \hline
  DJI & $4900$ & $0$ & $2597$ & $0$ & $9110$\\
  FutabaT14 & $5701$ & $110$ & $6208$ & $90$ & $9076$ \\
  FutabaT7 & $5086$ & $340$ & $6305$ & $180$ & $9226$ \\
  Noise & $2597$ & $560$ & $6358$ &  & \\
  Taranis & $5094$ & $670$ & $5942$ &  & \\
  Turnigy & $4032$ &  &  &  & \\
  \hline
\end{tabular}
\label{tab:field_test_frequencies}
\end{table}

% \begin{table}
% \centering
% \caption{Number of spectrograms for each class in the field measurements.}
% \label{tab:field_test_frequency1}
% %\setlength{\tabcolsep}{3pt}
% \begin{tabular}{ll}
% \hline
% \textbf{Class} & \textbf{Frequency} \\
% \hline
% DJI & 4900\\
% FutabaT14 & 5701\\
% FutabaT7 & 5086\\
% Noise & 2597 \\
% Taranis & 5094\\
% Turnigy & 4032\\
% \hline
% \end{tabular}
% \end{table}

% \begin{table}
% \centering
% \caption{Number of spectrograms for each distance whereas all spectrograms at 0~m are recorded without an active sender and thus labelled as ``Noise''.}
% \label{tab:field_test_frequency2}
% %\setlength{\tabcolsep}{3pt}
% \begin{tabular}{ll}
% \hline
% \textbf{Distance [m]} & \textbf{Frequency} \\
% \hline
% 0 & 2597\\
% 110 & 6208\\
% 340 & 6305\\
% 560 & 6358 \\
% 670 & 5942\\
% \hline
% \end{tabular}
% \end{table}

% \begin{table}
% \centering
% \caption{Number of spectrograms for each direction, given in angle relative to the sender.}
% \label{tab:field_test_frequency3}
% %\setlength{\tabcolsep}{3pt}
% \begin{tabular}{ll}
% \hline
% \textbf{Distance [m]} & \textbf{Frequency} \\
% \hline
% 0 & 9110\\
% 90 & 9074\\
% 180 & 9226\\
% \hline
% \end{tabular}
% \end{table}

%%%%%%%%%%%%%%%%%%%%%%%%%%%%%%%%%%%%%%%%%%
\section{Results}
\label{sec:results}
\subsection{Classification performance in the Development Dataset}
\label{sec:results_acc_dev_set}
Table~\ref{tab:acc_dev_set_multiclass} shows the general mean $\pm$ standard deviation of accuracy and balanced accuracy on the test data of the development dataset (cf.\ Sec.~\ref{sec:dev_dataset}), obtained in the $5$-fold cross-validation of the different models. 

There is no meaningful difference in performance between the models, even when the model complexity increases from VGG11\_BN to VGG19\_BN. The number of epochs for training (\#epochs) shows when the highest balanced accuracy was reached on the validation set. It can be seen that the least complex model, VGG11\_BN, required the least number of epochs compared to the more complex models. However, the resulting classification performance is the same.

\begin{table}
\centering
\caption{Mean $\pm$ standard deviation of the accuracy (Acc.) and the balanced accuracy (balanced Acc.) obtained in $5$-fold cross-validation of the different models on the test data of the development dataset. An indication of the model training time is given with the mean $\pm$ standard deviation of the number of training epochs (\#epochs), i.e.\ when the highest balanced accuracy on the validation set was reached. The number of trainable parameters (\#params) indicates the complexity of the model.}
\setlength{\tabcolsep}{3pt}
\begin{tabular}{cccccl}
  \hline
  \textbf{Model} & \textbf{Acc.} & \textbf{balanced Acc.} & \textbf{\#epochs} & \textbf{\#params}\\
  \hline
    VGG11\_BN & $0.944 \pm 0.005$ & $0.932 \pm 0.002$ & $66.4 \pm 24.4$ & $9.36 \cdot 10^6$\\
    VGG13\_BN & $0.947 \pm 0.003$ & $0.935 \pm 0.003$ & $138.6 \pm 46.4$ & $9.54 \cdot 10^6$ \\
    VGG16\_BN & $0.947 \pm 0.006$ & $0.937 \pm 0.005$ & $101.8 \pm 41.5$ & $14.86 \cdot 10^6$ \\
    VGG19\_BN & $0.952 \pm 0.006$ & $0.939 \pm 0.008$ & $98.2 \pm 45.8$ & $20.17 \cdot 10^6$ \\
\hline
\end{tabular}
\label{tab:acc_dev_set_multiclass}
\end{table}

\begin{figure}
\centering
\includegraphics[width=0.49\textwidth]%{figures/balanced_acc_over_SNR_mean.png}
{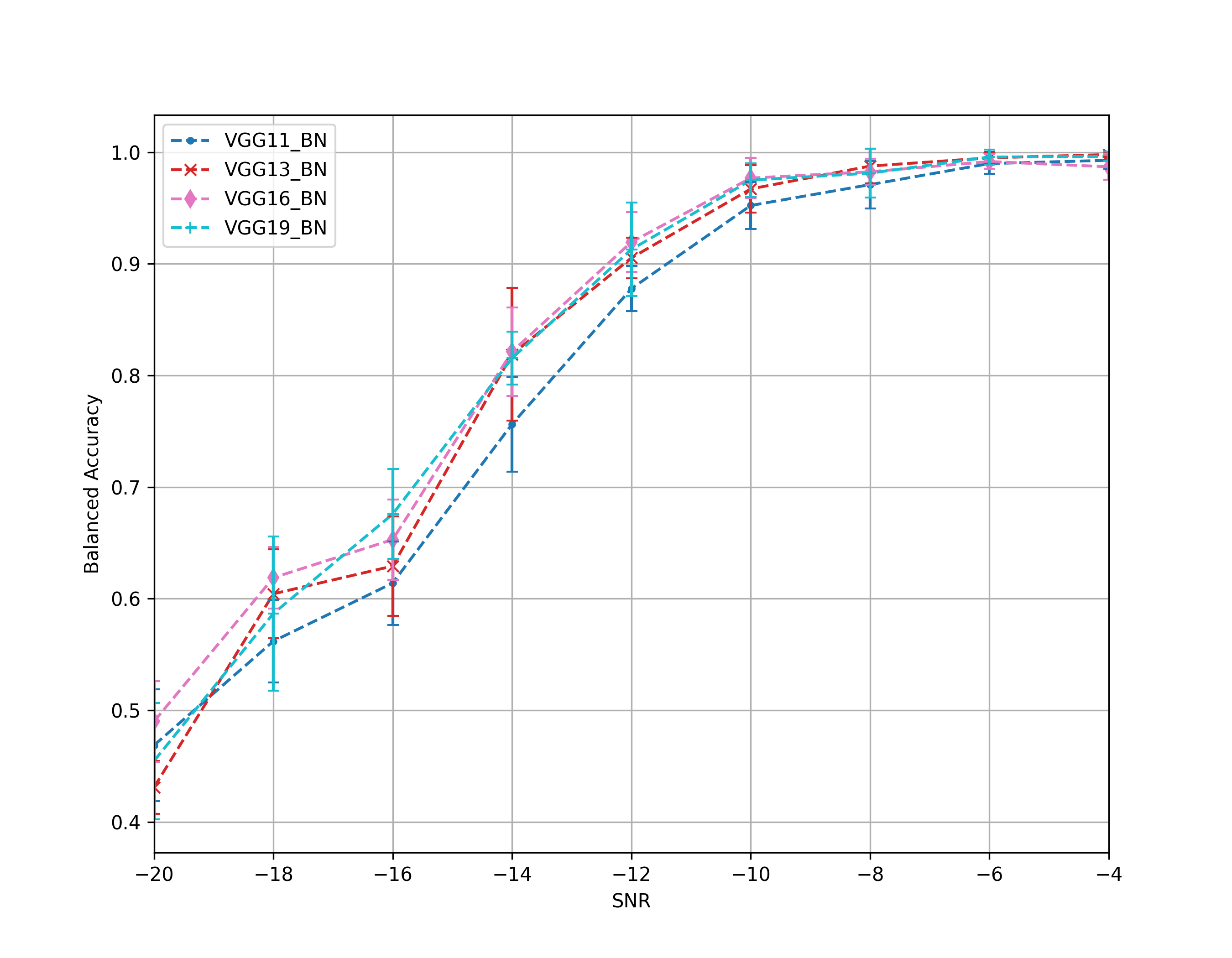}

\caption{Mean balanced accuracy $\pm$ std obtained in the $5$-fold cross-validation of the different models on the test set of the development dataset over the \glspl{snr} levels.}
\label{fig:balanced_acc_over_SNR}
\end{figure}

Figure~\ref{fig:balanced_acc_over_SNR} shows the resulting $5$-fold mean balanced accuracy over \glspl{snr} $\in [-20, -4]\,$dB in $2$\,dB steps. Note that we do not show the results for \glspl{snr} $> -4$dB, as those are simply saturated at 100$\%$ balanced accuracy.

In general, we observe a drastic degradation in performance from $-12$\,dB down to near chance level at $-20$\,dB.%Furthermore, there is no notable difference between the model architectures for the general mean (cf.\ Tab.~\ref{tab:acc_dev_set_multiclass}) .

\begin{figure}
\centering
\includegraphics[width=0.48\textwidth]{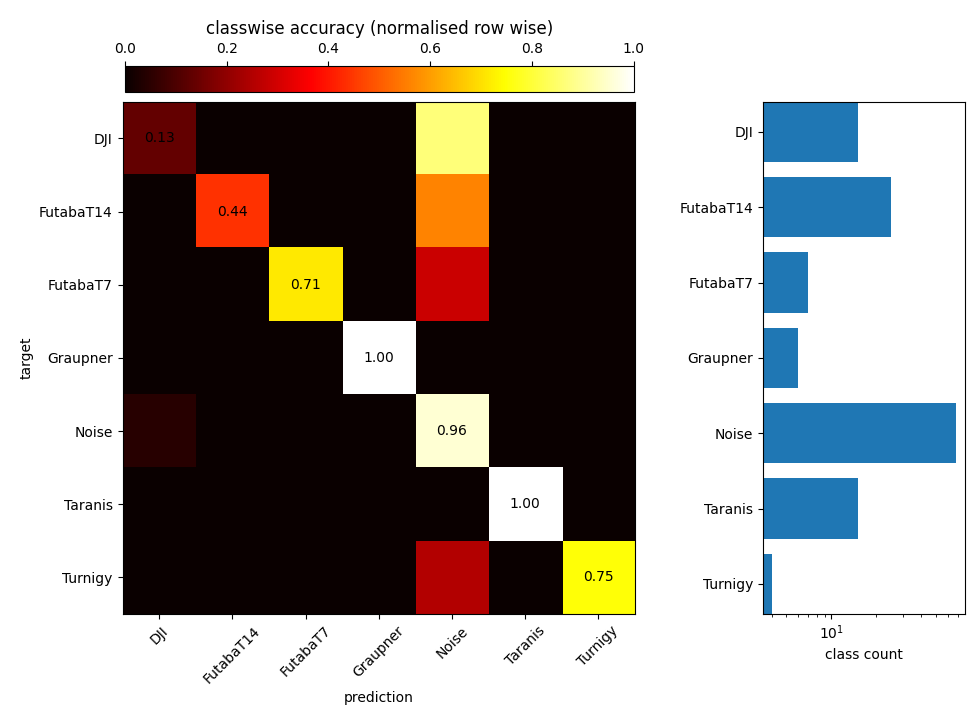}
\caption{Confusion matrix of the outputs of the VGG11\_BN model on a single fold for the samples at $-14$\,dB \gls{snr} from the test data. The average balanced accuracy is $0.71$.}
\label{fig:cm_vgg11_spec_SNR-14}
\end{figure}

The vast majority of misclassifications occurred between noise and drones and not between different types of drones. Figure~\ref{fig:cm_vgg11_spec_SNR-14} illustrates this fact. It shows the confusion matrix for the VGGG11\_BN model for a single validation on the test data for the samples with $-14$\,dB \gls{snr}.

\subsection{Embedding Space Visualisation}
\label{sec:results_embeddings}
\begin{figure}
\centering
\includegraphics[width=0.48\textwidth]{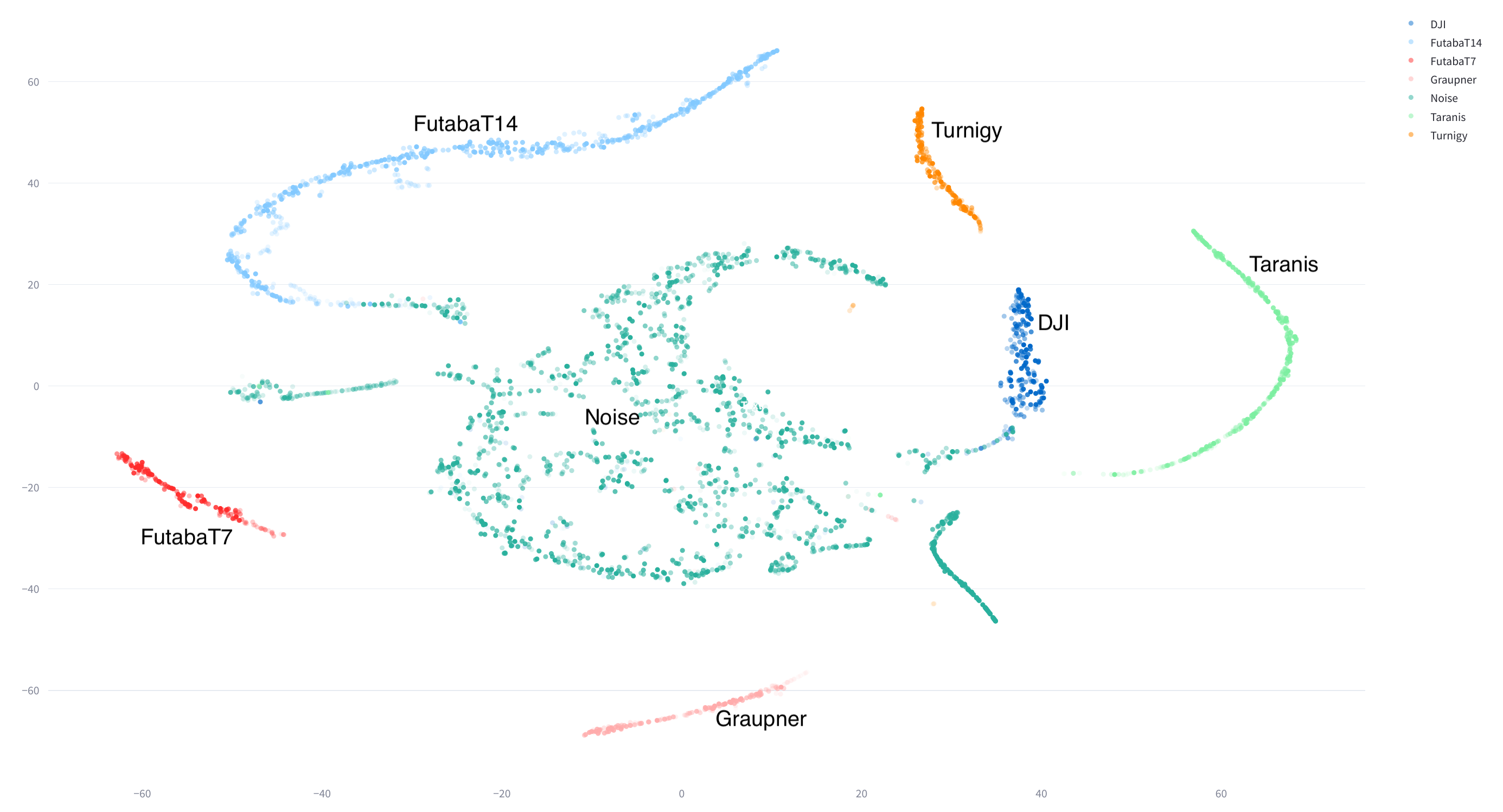}
\caption{2D \gls{tsne} visualisation of the VGG11\_BN embeddings of $3549$ test samples from the development dataset. The hyperparameters for \gls{tsne} were: metric ``eucleidean'', number of iterations $1000$, perplexity $30$ and method for gradient approximation ``barnes\_hut''}
\label{fig:vgg11_BN_embeddings}
\end{figure}

Figure~\ref{fig:vgg11_BN_embeddings} shows the 2D \gls{tsne} visualisation of the VGG11\_BN embeddings of $3549$ test samples from the development dataset. Each class forms a separate cluster. While the different drone signal clusters are rather small and dense, the noise cluster takes up most of the embedding space and even forms several sub-clusters. This is most likely due to the variety of the signals used in the noise class, i.e.\ Bluetooth and Wi-Fi signals plus Gaussian noise. It can also be seen that the DJI and FutabaT14 classes are more difficult to separate from noise than other classes (cf.~Fig.~\ref{fig:cm_vgg11_spec_SNR-14}) as they tend to overlap at the clusters' edges.

We used \gls{tsne} for dimensionality reduction because of its ability to preserve local structure within the high-dimensional embedding space. Furthermore, \gls{tsne} has been widely adopted in the \gls{ml} community and has a well-established track record for high-dimensional data visualisation. However, it is sensitive to hyperparameters such as perplexity and requires some tuning, i.e.\ different parameters can lead to considerable different results.

It can be argued that \gls{umap} would be a better choice due to its balanced preservation of local and global structure together with its robustness to hyperparameters. Therefore, we created a web application\footnote{\url{https://visvgg11bndronerfembeddings.streamlit.app}} that allows users to test and compare both approaches with different hyperparameters.

\subsection{Classification performance in the Field Test}
\label{sec:results_field_test}
For each model architecture, we performed $5$-fold cross-validation on the development dataset (cf.\ Sec.~\ref{sec:model_training}), resulting in five trained models per architecture. Thus, we also evaluated all five trained models on the field test data. We report the balanced accuracy $\pm$ standard deviation for each model architecture for the complete field test dataset averaged over all directions and distances in Tab.~\ref{tab:field_test_acc}.

\begin{table}
\caption{Mean $\pm$ standard deviation of the balanced accuracy (balanced Acc.) of the complete field test recordings for the different models.}
\centering
\begin{tabular}{cc}
\hline
\textbf{Model} & \textbf{balanced Acc.}  \\
\hline
VGG11\_BN & $0.792\pm0.022$ \\
VGG13\_BN & $0.807\pm0.011$ \\
VGG16\_BN & $0.811\pm0.009$ \\
VGG19\_BN & $0.806\pm0.016$ \\
\hline
\label{tab:field_test_acc}
\end{tabular}
\end{table}

As observed on the development dataset (cf.\ Tab.~\ref{tab:acc_dev_set_multiclass}), there is no meaningful difference in performance between the model architectures. We therefore focus on VGG11\_BN, the simplest model trained, in the more detailed analysis of the field test results.

A live system should trigger an alarm when a drone is present. Therefore, the question of whether the signal is from a drone at all is more important than predicting the correct type of drone. Therefore, we also evaluated the models in terms of a binary problem with two classes ``Drone'' (for all six classes of drones in the development dataset) and ``Noise''.

Table~\ref{tab:field_test_acc_type_direction} shows that the accuracies were highly depend on the class. Our models generalise well to the drones in the dataset, with the exception of the DJI. The dependence on direction is not as strong as expected. Orienting the antenna 180$^\circ$ away from the transmitter reduces the signal power by about $20$~dB, resulting  in lower \gls{snr} and lower classification accuracy. However, as the transmitters were still quite close to the antenna, the effect is not pronounced. As we have seen on the development dataset in Fig.~\ref{fig:balanced_acc_over_SNR}, there is a clear drop in accuracy once the \gls{snr} is below $-12$~dB. Apparently we were still above this threshold, regardless of the direction of the antenna.

\begin{table}
\caption{Mean balanced accuracy $\pm$ standard deviation of the VGG11\_BN models on the field test recordings for the different classes for each direction ($0^\circ$, $90^\circ$ and $180^\circ$). The upper part shows the accuracies for the classification problem (seven classes) and the lower part the accuracies for the detection problem ``Drone'' or ``Noise''}
\centering
\begin{tabular}{cccc}
\hline
\textbf{Class} & 
% \textbf{Frequency} & 
\textbf{$0^\circ$} & \textbf{$90^\circ$} & \textbf{$180^\circ$} \\
\hline
DJI	& $0.623\pm0.080$	& $0.624\pm0.035$ &	$0.540\pm0.051$\\
FutabaT14 & $0.716\pm0.101$ & $0.984\pm0.011$ &	$0.911\pm0.042$ \\
FutabaT7 & $0.724\pm0.041$	&$0.737\pm0.034$ &	$0.698\pm0.059$\\
Noise & $0.554\pm0.038$ &	$0.924\pm0.026$ &	$0.833\pm0.068$\\
Taranis	& $0.936\pm0.008$	& $0.879\pm0.002$	& $0.858\pm0.002$\\
Turnigy	& $0.899\pm0.129$ &	$0.958\pm0.027$ &	$0.962\pm0.024$\\
\hline\
Drone & $0.958\pm0.011$ &	$0.859\pm0.014$ & 	$0.847\pm0.017$\\
Noise & $0.554\pm0.038$ &	$0.924\pm0.026$ &	$0.833\pm0.068$\\
\hline
\label{tab:field_test_acc_type_direction}
\end{tabular}
\end{table}

What may be surprising is the low accuracy on the signals with no active transmitter, labelled as ``Noise'', in the direction of the lake ($0^\circ$). Given the uncontrolled nature of a field test, it could well be that there a drone was actually flying on the other side of the $2.3$~km wide lake. This could explain the false positives we observed in that direction.

Table~\ref{tab:field_test_acc_distance} shows the average balanced accuracy of the VGG11\_BN models on the field test data collected at different distances for each antenna direction. There is a slight decrease in accuracy with distance. However, the longest distance of $670$\,m appears to be too short to be a problem for the system. Unfortunately, this was the longest distance within line-of-sight that could be recorded at this location. 

\begin{table}
\caption{Mean balanced accuracy $\pm$ standard deviation of the VGG11\_BN models on the field test data with active transmitters collected at different distances for each antenna direction ($0^\circ$, $90^\circ$ and $180^\circ$). The upper part shows the accuracies for the classification problem (seven classes) and the lower part the accuracies for the detection problem ``Drone'' or ``Noise''}
\centering
\begin{tabular}{cccc}
\hline
\textbf{Classification}\\
Distance (m) & \textbf{$0^\circ$} & \textbf{$90^\circ$} & \textbf{$180^\circ$} \\
110 & $0.852\pm0.044$ & $0.838\pm0.024$ & $0.786\pm0.044$ \\
340 & $0.815\pm0.078$ & $0.916\pm0.017$ & $0.828\pm0.031$ \\
560 & $0.730\pm0.063$	& $0.796\pm0.007$ & $0.764\pm0.017$ \\
670 & $0.708\pm0.068$	& $0.805\pm0.008$ & $0.777\pm0.007$ \\
\hline
\textbf{Detection}\\
Distance (m) & \textbf{$0^\circ$} & \textbf{$90^\circ$} & \textbf{$180^\circ$} \\
110 & $0.955\pm0.007$ & $0.851\pm0.025$ & $0.849\pm0.018$ \\
340 & $0.986\pm0.008$ & $0.940\pm0.010$ & $0.913\pm0.018$ \\
560 & $0.960\pm0.011$ & $0.820\pm0.013$ & $0.807\pm0.022$ \\
670 & $0.925\pm0.021$ & $0.826\pm0.013$ & $0.823\pm0.014$ \\
\hline
\label{tab:field_test_acc_distance}
\end{tabular}
\end{table}

\begin{figure}[!t]
\centering
\includegraphics[width=0.48\textwidth]{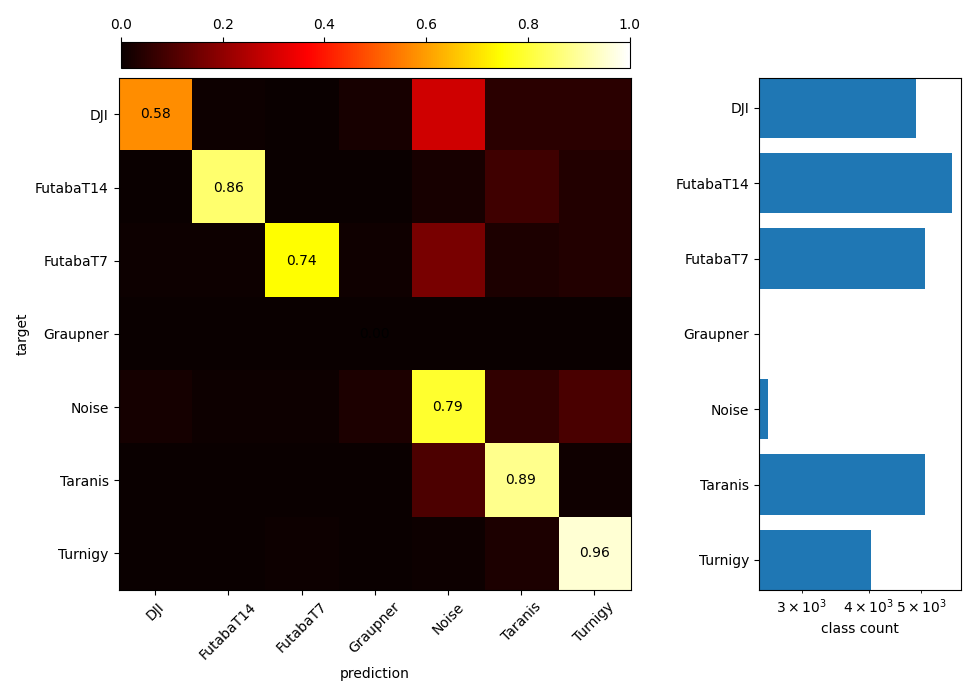}
\caption{Confusion matrix of the outputs of the VGG11\_BN model on a single fold for the samples from the field test data. The average balanced accuracy is $0.80$.}
\label{fig:cm_vgg11_field_test}
\end{figure}

Figure~\ref{fig:cm_vgg11_field_test} shows the confusion matrix for the outputs of the VGG11\_BN model of a single fold on the field test data. As with the development dataset (cf.\ Fig.~\ref{fig:cm_vgg11_spec_SNR-14}), most of the confusion is between noise and drones rather than between different types of drones. 

%%%%%%%%%%%%%%%%%%%%%%%%%%%%%%%%%%%%%%%%%%
\section{Discussion}
\label{sec:discussion}
We were able to show that a standard \gls{cnn}, trained on drone \gls{rf} signals recorded in a controlled laboratory environment and artificially augmented with noise, generalised well to the more challenging conditions of a real-world field test. 

The drone detection system consisted of rather simple and low budget hardware (consumer grade notebook with GPU + \gls{sdr}). Recording parameters such as sampling frequency, length of input vectors, etc. were set to enable real-time detection with the limited amount of memory and computing power. This means that data acquisition, pre-processing and model inference did not take longer than the signal being processed ($\approx 74.9$\,ms per sample in our case).  

Obviously, the \gls{vgg} models were able to learn the relevant features for the drone classification from the complex spectrograms of the \gls{rf} signal. In this respect, we did not find any advantage for the use of more complex models, such as VGG19\_BN, over the least complex model, VGG11\_BN (cf.\ Tabs.~\ref{tab:acc_dev_set_multiclass} and \ref{tab:field_test_acc}).

Furthermore, we have seen that the misclassifications mainly occur between the noise class and the drones, and not between the different drones themselves (cf.\ Figs.~\ref{fig:cm_vgg11_spec_SNR-14} and \ref{fig:cm_vgg11_field_test}). This is particularly relevant for the application of drone detection systems in security sensitive areas. The first priority is to detect any kind of \gls{uav}, regardless of its type. 

Based on our experience and results, we see the following limitations of our work. The \gls{rf} signals of drones were recorded in a controlled laboratory environment and augmented with WiFi/Bluetooth/Gaussian noise (cf.~Sec.~\ref{sec:material}). While the laboratory environment was necessary for the initial development and evaluation of the model, these signals do not fully replicate the complexities encountered in real-world scenarios. None the less, the field test showed that the models can be used and work reliably (cf.\ Tab.~\ref{tab:field_test_acc_type_direction}) in real world conditions. However, it is the nature of a field test that the level of interference from WiFi/Bluetooth noise and the possible presence of other drones cannot be fully controlled. Furthermore, due to the limited space/distance between the transmitter and receiver in our field test setup, we were not able to clearly demonstrate the effect of free space attenuation on detection performance (cf.\ Tab.~\ref{tab:field_test_acc_distance}).

Regarding the use of simple \glspl{cnn} as classifiers, it is not possible to reliably predict whether multiple transmitters are present. In that case, an object detection approach on the spectrogams could provide a more fine-grained prediction, see for example the works \cite{Prasad2020, Basak2022} and \cite{Zhao2024}. Nevertheless, the current approach will still detect a drone if one or more are present.

We have only tested a limited set of \gls{vgg} architectures. It remains to be seen whether more recent architectures, such as the pre-trained Vision Transformer \cite{dosovitskiy2020vit}, generalise as well or better. Further, with the deployment of the model on limited computational resources in mind, the use of more efficient and less complex models, such as GoogLeNet \cite{Szegedy2015} and ShuffleNet \cite{Zhang2018} should be investigated. Research could aim for a balance between detection accuracy and the computational demands of real-time processing. By doing so, the deployment of a robust, low-latency drone detection system in dynamic and resource-constrained settings could become a practical reality.

Another issue to consider is the occurrence of unknown drones, i.e.\ drones that are not part of the train set. Examining the embedding space (cf.\ \ref{sec:results_embeddings}) gives a first idea of whether a signal is clearly part of a known dense drone cluster or rather falls into the larger, less dense, noise cluster. We believe that a combination of an unsupervised deep autoencoder approach \cite{Lu2022, Zhou2023} with an additional classification part (cf.\ \cite{Pintelas2021}) would allow, first, to provide a stable classification of known samples and, second, to indicate whether a sample is known or rather an anomaly. This could be accompanied by additional field test to further access the model's generalisation ability to drones it was not trained on.

Finally, the current lack of standardised benchmarks makes it challenging to compare the performance of various models and systems objectively. We hope that our development dataset will inspire others to further optimise the model side of the problem and perhaps find a model architecture with better performance and/or efficiency.

\bibliographystyle{IEEEtran}
\bibliography{bibliography}

% Generated by IEEEtran.bst, version: 1.14 (2015/08/26)
\begin{thebibliography}{10}
\providecommand{\url}[1]{#1}
\csname url@samestyle\endcsname
\providecommand{\newblock}{\relax}
\providecommand{\bibinfo}[2]{#2}
\providecommand{\BIBentrySTDinterwordspacing}{\spaceskip=0pt\relax}
\providecommand{\BIBentryALTinterwordstretchfactor}{4}
\providecommand{\BIBentryALTinterwordspacing}{\spaceskip=\fontdimen2\font plus
\BIBentryALTinterwordstretchfactor\fontdimen3\font minus \fontdimen4\font\relax}
\providecommand{\BIBforeignlanguage}[2]{{%
\expandafter\ifx\csname l@#1\endcsname\relax
\typeout{** WARNING: IEEEtran.bst: No hyphenation pattern has been}%
\typeout{** loaded for the language `#1'. Using the pattern for}%
\typeout{** the default language instead.}%
\else
\language=\csname l@#1\endcsname
\fi
#2}}
\providecommand{\BIBdecl}{\relax}
\BIBdecl

\bibitem{Al-lQubaydhi2024}
\BIBentryALTinterwordspacing
N.~Al-lQubaydhi, A.~Alenezi, T.~Alanazi, A.~Senyor, N.~Alanezi, B.~Alotaibi, M.~Alotaibi, A.~Razaque, and S.~Hariri, ``Deep learning for unmanned aerial vehicles detection: A review,'' \emph{Computer Science Review}, vol.~51, p. 100614, 2 2024. [Online]. Available: \url{https://linkinghub.elsevier.com/retrieve/pii/S1574013723000813}
\BIBentrySTDinterwordspacing

\bibitem{Rahman2024}
\BIBentryALTinterwordspacing
M.~H. Rahman, M.~A.~S. Sejan, M.~A. Aziz, R.~Tabassum, J.-I. Baik, and H.-K. Song, ``A comprehensive survey of unmanned aerial vehicles detection and classification using machine learning approach: Challenges, solutions, and future directions,'' \emph{Remote Sensing}, vol.~16, p. 879, 3 2024. [Online]. Available: \url{https://www.mdpi.com/2072-4292/16/5/879}
\BIBentrySTDinterwordspacing

\bibitem{Allahham2019_DroneRF_dataset}
\BIBentryALTinterwordspacing
M.~S. Allahham, M.~F. Al-Sa'd, A.~Al-Ali, A.~Mohamed, T.~Khattab, and A.~Erbad, ``Dronerf dataset: A dataset of drones for rf-based detection, classification and identification,'' \emph{Data in Brief}, vol.~26, p. 104313, 10 2019. [Online]. Available: \url{https://linkinghub.elsevier.com/retrieve/pii/S2352340919306675}
\BIBentrySTDinterwordspacing

\bibitem{Al-Sad2019_DroneRF_dataset_article}
M.~F. Al-Sa'd, A.~Al-Ali, A.~Mohamed, T.~Khattab, and A.~Erbad, ``Rf-based drone detection and identification using deep learning approaches: An initiative towards a large open source drone database,'' \emph{Future Generation Computer Systems}, vol. 100, pp. 86--97, 11 2019.

\bibitem{Swinney2020}
C.~J. Swinney and J.~C. Woods, ``Unmanned aerial vehicle flight mode classification using convolutional neural network and transfer learning,'' in \emph{2020 16th International Computer Engineering Conference (ICENCO)}, 2020, pp. 83--87.

\bibitem{Zhang2021}
Y.~Zhang, ``Rf-based drone detection using machine learning,'' in \emph{2021 2nd International Conference on Computing and Data Science (CDS)}, 2021, pp. 425--428.

\bibitem{Ge2021}
C.~Ge, S.~Yang, W.~Sun, Y.~Luo, and C.~Luo, ``For rf signal-based uav states recognition, is pre-processing still important at the era of deep learning?'' in \emph{2021 7th International Conference on Computer and Communications (ICCC)}, 2021, pp. 2292--2296.

\bibitem{Xue2024}
\BIBentryALTinterwordspacing
Y.~Xue, Y.~Chang, Y.~Zhang, J.~Sun, Z.~Ji, H.~Li, Y.~Peng, and J.~Zuo, ``Uav signal recognition of heterogeneous integrated knn based on genetic algorithm,'' \emph{Telecommunication Systems}, vol.~85, pp. 591--599, 4 2024. [Online]. Available: \url{https://link.springer.com/10.1007/s11235-023-01099-x}
\BIBentrySTDinterwordspacing

\bibitem{AlKhonaini2024}
\BIBentryALTinterwordspacing
A.~AlKhonaini, T.~Sheltami, A.~Mahmoud, and M.~Imam, ``Uav detection using reinforcement learning,'' \emph{Sensors}, vol.~24, no.~6, 2024. [Online]. Available: \url{https://www.mdpi.com/1424-8220/24/6/1870}
\BIBentrySTDinterwordspacing

\bibitem{Ezuma2020}
\BIBentryALTinterwordspacing
M.~Ezuma, F.~Erden, C.~K. Anjinappa, O.~Ozdemir, and I.~Guvenc, ``Detection and classification of uavs using rf fingerprints in the presence of wi-fi and bluetooth interference,'' \emph{IEEE Open Journal of the Communications Society}, vol.~1, pp. 60--76, 2020. [Online]. Available: \url{https://ieeexplore.ieee.org/document/8913640/}
\BIBentrySTDinterwordspacing

\bibitem{Ezuma2020data_drone_remote_controller}
\BIBentryALTinterwordspacing
------, ``Drone remote controller rf signal dataset,'' 2020. [Online]. Available: \url{https://dx.doi.org/10.21227/ss99-8d56}
\BIBentrySTDinterwordspacing

\bibitem{Ozturk2021}
\BIBentryALTinterwordspacing
E.~Ozturk, F.~Erden, and I.~Guvenc, ``Rf-based low-snr classification of uavs using convolutional neural networks,'' \emph{ITU Journal on Future and Evolving Technologies}, vol.~2, pp. 39--52, 7 2021. [Online]. Available: \url{https://www.itu.int/pub/S-JNL-VOL2.ISSUE5-2021-A04}
\BIBentrySTDinterwordspacing

\bibitem{Sweeny2021data_drone_detect}
\BIBentryALTinterwordspacing
C.~J. Swinney and J.~C. Woods, ``Dronedetect dataset: A radio frequency dataset of unmanned aerial system (uas) signals for machine learning detection \& classification,'' 2021. [Online]. Available: \url{https://dx.doi.org/10.21227/5jjj-1m32}
\BIBentrySTDinterwordspacing

\bibitem{Swinney2021}
------, ``Rf detection and classification of unmanned aerial vehicles in environments with wireless interference,'' in \emph{2021 International Conference on Unmanned Aircraft Systems (ICUAS)}, 2021, pp. 1494--1498.

\bibitem{Kunze2022_drone_cnn_raw_iq_data}
\BIBentryALTinterwordspacing
S.~Kunze and B.~Saha, ``Drone classification with a convolutional neural network applied to raw iq data,'' in \emph{2022 3rd URSI Atlantic and Asia Pacific Radio Science Meeting (AT-AP-RASC)}, May 2022, pp. 1--4. [Online]. Available: \url{https://ieeexplore.ieee.org/document/9814170/}
\BIBentrySTDinterwordspacing

\bibitem{Medaiyese2022_data_cardRF}
\BIBentryALTinterwordspacing
O.~Medaiyese, M.~Ezuma, A.~Lauf, and A.~Adeniran, ``Cardinal rf (cardrf): An outdoor uav/uas/drone rf signals with bluetooth and wifi signals dataset,'' 2022. [Online]. Available: \url{https://dx.doi.org/10.21227/1xp7-ge95}
\BIBentrySTDinterwordspacing

\bibitem{Medaiyese2021_semi_sup_framework}
O.~O. Medaiyese, M.~Ezuma, A.~P. Lauf, and A.~A. Adeniran, ``Hierarchical learning framework for uav detection and identification,'' \emph{IEEE Journal of Radio Frequency Identification}, vol.~6, pp. 176--188, 2022.

\bibitem{Medaiyese2022}
\BIBentryALTinterwordspacing
O.~O. Medaiyese, M.~Ezuma, A.~P. Lauf, and I.~Guvenc, ``Wavelet transform analytics for rf-based uav detection and identification system using machine learning,'' \emph{Pervasive and Mobile Computing}, vol.~82, p. 101569, 6 2022. [Online]. Available: \url{https://linkinghub.elsevier.com/retrieve/pii/S1574119222000219}
\BIBentrySTDinterwordspacing

\bibitem{Iandola2016_squeezenet}
\BIBentryALTinterwordspacing
F.~N. Iandola, M.~W. Moskewicz, K.~Ashraf, S.~Han, W.~J. Dally, and K.~Keutzer, ``Squeezenet: Alexnet-level accuracy with 50x fewer parameters and {\textless}1mb model size,'' \emph{CoRR}, vol. abs/1602.07360, 2016. [Online]. Available: \url{http://arxiv.org/abs/1602.07360}
\BIBentrySTDinterwordspacing

\bibitem{Gluege2023}
S.~Glüge., M.~Nyfeler., N.~Ramagnano., C.~Horn., and C.~Schüpbach., ``Robust drone detection and classification from radio frequency signals using convolutional neural networks,'' in \emph{Proceedings of the 15th International Joint Conference on Computational Intelligence - NCTA}, INSTICC.\hskip 1em plus 0.5em minus 0.4em\relax SciTePress, 2023, pp. 496--504.

\bibitem{Zhao2024}
\BIBentryALTinterwordspacing
R.~Zhao, T.~Li, Y.~Li, Y.~Ruan, and R.~Zhang, ``Anchor-free multi-uav detection and classification using spectrogram,'' \emph{IEEE Internet of Things Journal}, vol.~11, pp. 5259--5272, 2 2024. [Online]. Available: \url{https://ieeexplore.ieee.org/document/10221859/}
\BIBentrySTDinterwordspacing

\bibitem{Sun2017}
C.~Sun, A.~Shrivastava, S.~Singh, and A.~Gupta, ``Revisiting unreasonable effectiveness of data in deep learning era,'' in \emph{2017 IEEE International Conference on Computer Vision (ICCV)}, 2017, pp. 843--852.

\bibitem{2020SciPy-NMeth}
P.~Virtanen, R.~Gommers, T.~E. Oliphant, M.~Haberland, T.~Reddy, D.~Cournapeau, E.~Burovski, P.~Peterson, W.~Weckesser, J.~Bright, S.~J. {van der Walt}, M.~Brett, J.~Wilson, K.~J. Millman, N.~Mayorov, A.~R.~J. Nelson, E.~Jones, R.~Kern, E.~Larson, C.~J. Carey, {\.I}.~Polat, Y.~Feng, E.~W. Moore, J.~{VanderPlas}, D.~Laxalde, J.~Perktold, R.~Cimrman, I.~Henriksen, E.~A. Quintero, C.~R. Harris, A.~M. Archibald, A.~H. Ribeiro, F.~Pedregosa, P.~{van Mulbregt}, and {SciPy 1.0 Contributors}, ``{{SciPy} 1.0: Fundamental Algorithms for Scientific Computing in Python},'' \emph{Nature Methods}, vol.~17, pp. 261--272, 2020.

\bibitem{Simonyan2015_vgg}
\BIBentryALTinterwordspacing
K.~Simonyan and A.~Zisserman, ``Very deep convolutional networks for large-scale image recognition,'' in \emph{3rd International Conference on Learning Representations, {ICLR} 2015, San Diego, CA, USA, May 7-9, 2015, Conference Track Proceedings}, Y.~Bengio and Y.~LeCun, Eds., 2015. [Online]. Available: \url{http://arxiv.org/abs/1409.1556}
\BIBentrySTDinterwordspacing

\bibitem{Shin2016}
H.-C. Shin, H.~R. Roth, M.~Gao, L.~Lu, Z.~Xu, I.~Nogues, J.~Yao, D.~Mollura, and R.~M. Summers, ``Deep convolutional neural networks for computer-aided detection: Cnn architectures, dataset characteristics and transfer learning,'' \emph{IEEE Transactions on Medical Imaging}, vol.~35, no.~5, pp. 1285--1298, 2016.

\bibitem{Shaoqing2015}
S.~Ren, K.~He, R.~Girshick, and J.~Sun, ``Faster r-cnn: towards real-time object detection with region proposal networks,'' in \emph{Proceedings of the 28th International Conference on Neural Information Processing Systems - Volume 1}, ser. NIPS'15.\hskip 1em plus 0.5em minus 0.4em\relax Cambridge, MA, USA: MIT Press, 2015, p. 91–99.

\bibitem{Hershey2017}
S.~Hershey, S.~Chaudhuri, D.~P.~W. Ellis, J.~F. Gemmeke, A.~Jansen, R.~C. Moore, M.~Plakal, D.~Platt, R.~A. Saurous, B.~Seybold, M.~Slaney, R.~J. Weiss, and K.~Wilson, ``Cnn architectures for large-scale audio classification,'' in \emph{2017 IEEE International Conference on Acoustics, Speech and Signal Processing (ICASSP)}, 2017, pp. 131--135.

\bibitem{Ioffe2015_batchnorm}
S.~Ioffe and C.~Szegedy, ``Batch normalization: accelerating deep network training by reducing internal covariate shift,'' in \emph{Proceedings of the 32nd International Conference on International Conference on Machine Learning - Volume 37}, ser. ICML'15.\hskip 1em plus 0.5em minus 0.4em\relax JMLR.org, 2015, p. 448–456.

\bibitem{Paszke2019pytorch}
A.~Paszke, S.~Gross, F.~Massa, A.~Lerer, J.~Bradbury, G.~Chanan, T.~Killeen, Z.~Lin, N.~Gimelshein, L.~Antiga, A.~Desmaison, A.~Kopf, E.~Yang, Z.~DeVito, M.~Raison, A.~Tejani, S.~Chilamkurthy, B.~Steiner, L.~Fang, J.~Bai, and S.~Chintala, ``Pytorch: An imperative style, high-performance deep learning library,'' in \emph{Advances in Neural Information Processing Systems 32}, H.~Wallach, H.~Larochelle, A.~Beygelzimer, F.~d\textquotesingle Alch\'{e}-Buc, E.~Fox, and R.~Garnett, Eds.\hskip 1em plus 0.5em minus 0.4em\relax Curran Associates, Inc., 2019, pp. 8024--8035.

\bibitem{Diederik2015_adam_optimizer}
\BIBentryALTinterwordspacing
D.~P. Kingma and J.~Ba, ``Adam: {A} method for stochastic optimization,'' in \emph{3rd International Conference on Learning Representations, {ICLR} 2015, San Diego, CA, USA, May 7-9, 2015, Conference Track Proceedings}, Y.~Bengio and Y.~LeCun, Eds., 2015. [Online]. Available: \url{http://arxiv.org/abs/1412.6980}
\BIBentrySTDinterwordspacing

\bibitem{vandermaaten2008_tsne}
\BIBentryALTinterwordspacing
L.~van~der Maaten and G.~Hinton, ``Visualizing data using t-sne,'' \emph{Journal of Machine Learning Research}, vol.~9, no.~86, pp. 2579--2605, 2008. [Online]. Available: \url{http://jmlr.org/papers/v9/vandermaaten08a.html}
\BIBentrySTDinterwordspacing

\bibitem{McInnes2018_umap}
\BIBentryALTinterwordspacing
L.~McInnes, J.~Healy, N.~Saul, and L.~Großberger, ``Umap: Uniform manifold approximation and projection,'' \emph{Journal of Open Source Software}, vol.~3, no.~29, p. 861, 2018. [Online]. Available: \url{https://doi.org/10.21105/joss.00861}
\BIBentrySTDinterwordspacing

\bibitem{Prasad2020}
K.~N.~R. Surya Vara~Prasad and V.~K. Bhargava, ``A classification algorithm for blind uav detection in wideband rf systems,'' in \emph{2020 IEEE 92nd Vehicular Technology Conference (VTC2020-Fall)}, 2020, pp. 1--7.

\bibitem{Basak2022}
S.~Basak, S.~Rajendran, S.~Pollin, and B.~Scheers, ``Combined rf-based drone detection and classification,'' \emph{IEEE Transactions on Cognitive Communications and Networking}, vol.~8, no.~1, pp. 111--120, 2022.

\bibitem{dosovitskiy2020vit}
A.~Dosovitskiy, L.~Beyer, A.~Kolesnikov, D.~Weissenborn, X.~Zhai, T.~Unterthiner, M.~Dehghani, M.~Minderer, G.~Heigold, S.~Gelly, J.~Uszkoreit, and N.~Houlsby, ``An image is worth 16x16 words: Transformers for image recognition at scale,'' \emph{ICLR}, 2021.

\bibitem{Szegedy2015}
\BIBentryALTinterwordspacing
C.~Szegedy, W.~Liu, Y.~Jia, P.~Sermanet, S.~Reed, D.~Anguelov, D.~Erhan, V.~Vanhoucke, and A.~Rabinovich, ``Going deeper with convolutions,'' in \emph{2015 IEEE Conference on Computer Vision and Pattern Recognition (CVPR)}.\hskip 1em plus 0.5em minus 0.4em\relax Los Alamitos, CA, USA: IEEE Computer Society, jun 2015, pp. 1--9. [Online]. Available: \url{https://doi.ieeecomputersociety.org/10.1109/CVPR.2015.7298594}
\BIBentrySTDinterwordspacing

\bibitem{Zhang2018}
\BIBentryALTinterwordspacing
X.~Zhang, X.~Zhou, M.~Lin, and J.~Sun, ``Shufflenet: An extremely efficient convolutional neural network for mobile devices,'' in \emph{2018 IEEE/CVF Conference on Computer Vision and Pattern Recognition (CVPR)}.\hskip 1em plus 0.5em minus 0.4em\relax Los Alamitos, CA, USA: IEEE Computer Society, jun 2018, pp. 6848--6856. [Online]. Available: \url{https://doi.ieeecomputersociety.org/10.1109/CVPR.2018.00716}
\BIBentrySTDinterwordspacing

\bibitem{Lu2022}
\BIBentryALTinterwordspacing
S.~Lu and R.~Li, \emph{DAC–Deep Autoencoder-Based Clustering: A General Deep Learning Framework of Representation Learning}.\hskip 1em plus 0.5em minus 0.4em\relax Springer Science and Business Media Deutschland GmbH, 2 2022, vol. 294, pp. 205--216. [Online]. Available: \url{https://link.springer.com/10.1007/978-3-030-82193-7_13}
\BIBentrySTDinterwordspacing

\bibitem{Zhou2023}
\BIBentryALTinterwordspacing
H.~Zhou, J.~Bai, Y.~Wang, J.~Ren, X.~Yang, and L.~Jiao, ``Deep radio signal clustering with interpretability analysis based on saliency map,'' \emph{Digital Communications and Networks}, 1 2023. [Online]. Available: \url{https://linkinghub.elsevier.com/retrieve/pii/S2352864823000238}
\BIBentrySTDinterwordspacing

\bibitem{Pintelas2021}
\BIBentryALTinterwordspacing
E.~Pintelas, I.~E. Livieris, and P.~E. Pintelas, ``A convolutional autoencoder topology for classification in high-dimensional noisy image datasets,'' \emph{Sensors}, vol.~21, p. 7731, 11 2021. [Online]. Available: \url{https://www.mdpi.com/1424-8220/21/22/7731}
\BIBentrySTDinterwordspacing

\end{thebibliography}

\begin{IEEEbiography}[{\includegraphics[width=1in,height=1.25in,clip,keepaspectratio]{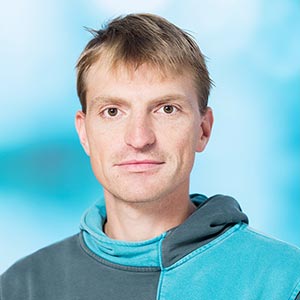}}]{Stefan Glüge}~is a senior researcher at the Institute for Computational Life Sciences (ICLS) at the Zurich University of Applied Sciences (ZHAW) in Switzerland. With over a decade of experience, Dr.~Glüge specializes in machine learning, particularly in developing learning algorithms for recurrent neural networks. He completed his PhD in implicit sequence learning in recurrent neural networks from Otto-von-Guericke University Magdeburg, where he also earned his Master’s degree in engineering with a focus on context-dependent learning for biological behavior modeling. Dr.~Glüge has made significant contributions to the field of applied machine learning and computer vision, participating in various Innosuisse/CTI projects. His research includes deep learning applications in medical imaging, automated monitoring systems, and pattern recognition. Notable projects involve AI for hematological disease classification, machine learning for equine reproductive monitoring, and explainable deep learning models for medical time series data.% His work has been widely published in scientific journals and conference proceedings, solidifying his reputation as a leader in his field.
% Shorter version
%Dr. Stefan Glüge is a senior researcher at the Institute for Computational Life Sciences (ICLS) at Zurich University of Applied Sciences (ZHAW). He holds a PhD in implicit sequence learning in recurrent neural networks from Otto-von-Guericke University Magdeburg. With over ten years of experience in machine learning and computer vision, Dr. Glüge's research focuses on AI applications in medical imaging, disease classification, and automated monitoring systems. His work is widely published and recognized in the scientific community.
\end{IEEEbiography}

\begin{IEEEbiography}[{\includegraphics[width=1in,height=1.25in,clip,keepaspectratio]{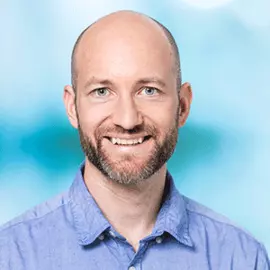}}]{Matthias Nyfeler}~Matthias Nyfeler is a senior lecturer at the Institute for Computational Life Sciences (ICLS) at the Zurich University of Applied Sciences (ZHAW) in Switzerland. He completed his PhD in theoretical physics on the numerical simulations of strongly correlated electron systems at the Albert Einstein Center for Fundamental Physics at the University of Bern, Switzerland. He teaches mathematical modelling, machine learning, statistics lectures and heads the Applied Computational Life Sciences Master’s programme and has been leading various projects on deep learning, drone radio signal detection, bioacoustics, signal processing, physical computing, and education.
\end{IEEEbiography}

\begin{IEEEbiography}[{\includegraphics[width=1in,height=1.25in,clip,keepaspectratio]{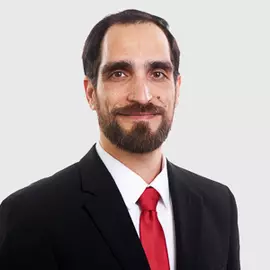}}]{Ahmad Aghaebrahimian}~is a research associate at the Institute for Computational Life Sciences (ICLS) at the Zurich University of Applied Sciences (ZHAW) in Switzerland. He is a computer scientist and linguist by training. He did his PhD in computer science in Charles University in Prague, majoring in Natural Language Processing (NLP) and Deep Learning. Throughout the last several years, he has been the PI and co-PI of several Innosuisse and DIZH projects in medical text analytics, food technology, supply chain monitoring, and bioinformatics. His areas of interest include NLP, Large Language Models, Neural-Symbolic Learning, Deep Learning, and semantic web.
\end{IEEEbiography}

\begin{IEEEbiography}[{\includegraphics[width=1in,height=1.25in,clip,keepaspectratio]{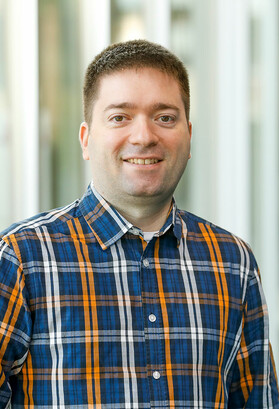}}]{Nicola Ramagnano}~Nicola Ramagnano received the diploma in electrical engineering (dipl. Ing. FH in Elektrotechnik) from the University of Applied Sciences Eastern Switzerland, Rapperswil, HSR. After a few years in industry, he is currently a research assistant at the Institute for Communication Systems (ICOM) of the Eastern Switzerland University of Applied Sciences (OST). His main areas of interest include signal processing, wireless communications, software defined radio (SDR), as well as RF and microwave electronics.
\end{IEEEbiography}

\begin{IEEEbiography}[{\includegraphics[width=1in,height=1.25in,clip,keepaspectratio]{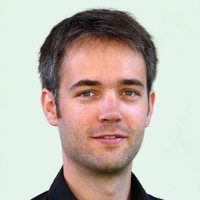}}]{Christof Schüpbach}~received the Ph.D.\ and M.Sc.\ degrees in theoretical particle physics from the Albert Einstein Center for Fundamental Physics, University of Bern, Bern, Switzerland. He is currently a Scientific Project Manager with the Communications, Specialised Service Networks and Protection group, armasuisse Science and Technology, within the Swiss Department of Defense. He leads the armasuisse research project on passive radar using civilian digital broadcasting transmitters, and works on projects in the fields of electronic warfare, timing and synchronisation, and software defined radio.
\end{IEEEbiography}

\end{document}